\documentclass[preprint,authoryear,12pt]{elsarticle}

\usepackage{epsfig}

\usepackage{amssymb}

\journal{Advances in Space Research}

\def\avg{\bar}
\newcommand{\dv}{\mbox{\boldmath $\nabla \cdot$}}
\newcommand{\vv}{\mbox{\boldmath $\vec{v}$}}
\newcommand{\bdot}{\mbox{\boldmath $\cdot$}}
\newcommand{\rhat}{\mbox{\boldmath $\vec{e}_r$}}
\newcommand{\ggrav}{\mbox{\boldmath $\vec{g}$}}
\newcommand{\pd}{\partial}
\newcommand{\del}{\mbox{\boldmath $\nabla$}}
\newcommand{\cross}{\mbox{\boldmath $\times$}}
\newcommand{\scrD}{\mbox{\boldmath ${\cal D}$}}
\def\scrR{\mathcal{R}}

\begin{document}

\begin{frontmatter}



\title{Velocity Amplitudes in Global Convection Simulations: The Role of the Prandtl Number and Near-Surface Driving}


\author[label1]{Bridget O'Mara}
\cortext[cor]{Corresponding author}
\ead{bomara@regis.edu}


\author[label2]{Mark S. Miesch\corref{cor}}
\ead{miesch@ucar.edu}

\author[label3]{Nicholas A. Featherstone}
\ead{nicholas.featherstone@colorado.edu}

\author[label2]{Kyle C. Augustson}
\ead{kyle.augustson@gmail.com}

\address[label1]{Physics Department, Regis University, 3333 Regis Blvd, Denver, CO, 80221, USA}
\address[label2]{High Altitude Observatory, National Center for Atmospheric Research, 3080 Center Green Dr., Boulder, CO 80301, USA}
\address[label3]{Dept.\ Applied Mathematics, Univ.\ of Colorado, Boulder, CO 80309, USA}

\begin{abstract}

Several lines of evidence suggest that the velocity amplitude in global simulations of solar convection, $U$, may be systematically over-estimated.  Motivated by these recent results, we explore the factors that determine $U$ and we consider how these might scale to solar parameter regimes.  To this end, we decrease the thermal diffusivity $\kappa$ along two paths in parameter space.  If the kinematic viscosity $\nu$ is decreased proportionally with $\kappa$ (fixing the Prandtl number $P_r = \nu/\kappa$), we find that $U$ increases but asymptotes toward a constant value, as found by \cite{feath16}.   However, if $\nu$ is held fixed while decreasing $\kappa$ (increasing $P_r$), we find that $U$ systematically decreases.  We attribute this to an enhancement of the thermal content of downflow plumes, which allows them to carry the solar luminosity with slower flow speeds.  We contrast this with the case of Rayleigh-B\'enard convection which is not subject to this luminosity constraint.  This dramatic difference in behavior for the two paths in parameter space (fixed $P_r$ or fixed $\nu$) persists whether the heat transport by unresolved, near-surface convection is modeled as a thermal conduction or as a fixed flux.  The results suggest that if solar convection can operate in a high-$P_r$ regime, then this might effectively limit the velocity amplitude.  Small-scale magnetism is a possible source of enhanced viscosity that may serve to achieve this high-$P_r$ regime.

\end{abstract}


\end{frontmatter}

\parindent=0.5 cm

\section{Introduction}\label{sec:intro}

Observations clearly indicate that the surface of the Sun is blanketed by roughly a million convection cells, each with a characteristic size of about 1.4 Mm and a typical lifetime of about 10-15 min.  This is the well-known phenomena of solar granulation and it represents the dominant scale of convection in the solar photosphere.  Granulation is well understood theoretically and modern numerical simulations are able to reproduce observations with striking fidelity \citep{nordl09}.  In fact, modeling solar granulation is arguably one of the most notable success stories in all of computational astrophysics.

However, there is far more to solar convection than solar granulation.  Granulation alone cannot account for the differential rotation and meridional circulation apparent from surface observations and helioseismic inversions.  Furthermore, it is these mean flows, together with the deeper, larger-scale convective motions that must maintain them, that are thought to be responsible for establishing the solar activity cycle \citep{miesc05,charb10}.

There are three recent research developments that suggest our current understanding of deep solar convection may be incomplete.  The first comes from the same numerical simulations of solar surface convection that do so well at capturing granulation.  When these are extended to larger horizontal and vertical scales, they tend to over-estimate the power at low spectral wavenumbers when compared to photospheric observations \citep{lord14}.  For the largest horizontal scales in the simulations, 196 Mm, the characteristic amplitude of convective velocities is at least three times larger than that inferred from correlation tracking of photospheric features, with the trend suggesting even larger discrepancies at larger scales.  Furthermore, the peak vertical velocities at the bottom of the computational domain, 49 Mm below the solar surface, can exceed several hundred m s$^{-1}$.  Though these surface convection simulations neglect rotation, such large convective velocities suggest a weak rotational influence.  

To demonstrate this, we can quantify the rotational influence by means of the Rossby number $R_o = U/(2 \Omega L)$ where $\Omega$ is some measure of the rotation rate and $U$ and $L$ are typical convective velocity and length scales.  If we take $\Omega = 2.7 \times 10^{-6}$ rad s$^{-1}$ $U \sim $ 200 m s$^{-1}$, and $L \sim H_\rho \sim $ 28 Mm, where $H_\rho$ is the density scale height, we obtain $R_o \sim 1.3$.  This appears to be at odds with the differential rotation of the Sun inferred from photospheric observations and helioseismic inversions, which reveal a $\sim$30\% decrease in angular velocity from equator to pole.  Such a substantial differential rotation implies efficient angular momentum transport by the convective Reynolds stress \citep{miesc05}.  This in turn requires the strong Coriolis-induced velocity correlations that are generally associated with $R_o \ll 1$.

This expectation is borne out by global simulations of solar and stellar convection, which show a dramatic change in the differential rotation profile near $R_o = 1$ \citep{guerr13,gasti14,fan14,kapyl14,feath15,karak15}.  For $R_o \ll 1$, differential rotation profiles are generally solar-like in the sense that the $\Omega$ gradient is equatorward, decreasing from equator to pole (though the $\Omega$ contours are often more cylindrically-aligned than in the Sun).  For $R_o \gtrsim 1$, differential rotation profiles are anti-solar, with a poleward $\Omega$ gradient.  This is attributed to the tendency for convective motions to conserve their angular momentum locally when the rotational influence is weak.

Global simulations of solar convection often fall in the anti-solar rotation regime, suggesting that they are over-estimating the Rossby number of the deep solar convection zone.  This contradiction is consistent with the velocity amplitudes of surface convection simulations described above but represents a second, independent recent research development that calls into question our current understanding of deep solar convection.  Current global simulations of solar convection often artificially reduce the Rossby number by increasing the rotation rate \citep{augus15,karak15} or by decreasing the convective luminosity, which in turn suppresses $U$ \citep{racin11,guerr13,fan14,hotta15a}.  

Recent results from local helioseismology also suggest that numerical and theoretical models may be over-estimating the velocity amplitudes of deep solar convection, though the evidence is mixed.  This is the third development referred to above.  \citet{hanas10,hanas12} searched for signatures of giant cells in helioseismic time-distance measurements and did not find a significant signal.  They placed an upper limit of less than 3 m s$^{-1}$ for each spherical harmonic degree $\ell$ at $r = 0.96R$ for convective motions with coherence times longer than 24 hours and $\ell < 60$, where $R$ is the solar radius.  When the coherence time was increased to 96 hours, they inferred amplitudes of less than 1 m s$^{-1}$ for each $\ell$.  When summed over all spherical harmonic modes $\ell < 60$, this corresponds to an upper limit on $U$ of less than 10 m s$^{-1}$.  This upper limit is more than an order of magnitude lower than the convective velocities typically found in global convection simulations and predicted from mixing-length theory.  Furthermore, it is hard to reconcile with helioseismic measurements of the differential rotation and meridional circulation \citep{miesc12b}.  More recent helioseismic estimates of the convective flow speed using ring-diagram analysis by \citet{greer15} suggest values that are closer to the theoretical models.  They find $U \sim$ 120 m s$^{-1}$ at $r \sim $ 0.96-0.97 $R$.  For a recent review of these and other results, see \citet{hanas16}.

Though each of these three recent developments is subject to its own uncertainties, they collectively suggest that there may be something fundamental about deep solar convection that we are missing.  In short, we may state the {\em \bf convective conundrum} as follows: The convective velocities required to transport the solar luminosity in global models of solar convection appear to be systematically larger than those required to maintain the solar differential rotation and those inferred from solar observations.

Even more to the point of this paper, the problem seems to get worse as the dissipation is decreased.  Due to computational constraints, the values of the kinematic viscosity $\nu$ and the thermal diffusivity $\kappa$ that are currently used in global convection simulations are orders of magnitude larger than the corresponding values of the actual solar plasma.  One would hope that the conundrum might resolve itself if $\nu$ and $\kappa$ could be drastically reduced, making the parameter regime closer to the Sun.  However, simulations have instead found that lowering $\nu$ and $\kappa$ can exacerbate the problem by further increasing the effective Rossby number \citep{guerr13,gasti14,feath15,karak15}. 

Motivated by this conundrum, we have performed idealized numerical experiments to investigate the nature of the convective driving in global simulations.  We focus in particular on significance of the Prandtl number, which is the ratio of viscous to thermal diffusion; $P_r = \nu/\kappa$.  We demonstrate that a high value of $P_r$ helps to limit $U$ as $\kappa$ is decreased and we argue that this may help to resolve the convective conundrum.  We also demonstrate that this conclusion is insensitive to the details of how the heat transport by unresolved near-surface convection is parameterized.

In section \ref{sec:Rayleigh} we describe the Rayleigh code, which we use to solve the anelastic fluid equations in a spherical shell, as well as the setup of the numerical experiments.  For these fundamental studies we neglect rotation.  We present the results of our numerical experiments in sections \ref{sec:Prandtl} and \ref{sec:fg}, which focus on the role of the Prandtl number and the upper boundary layer respectively.  We close with a discussion and summary of our results and their implications in sections \ref{sec:discussion} and \ref{sec:summary}.

\section{Numerical Experiments}\label{sec:Rayleigh}

\subsection{Numerical Model}\label{sec:model}

This study is based around a series of 3-D, nonlinear convection models run with the {\em Rayleigh} convection code.  {\em Rayleigh} is a modern, parallel code that has been designed to simulate thermal convection in spherical geometries at high resolution; see \citet[][hereafter FH16]{feath16} and a forthcoming code paper for further details.  

{\em Rayleigh} employs a pseudo-spectral approach, meaning that derivatives along each axis are calculated in the relevant spectral configuration, whereas   nonlinear terms are calculated after first transforming to physical space.  We represent the horizontal variation of all variables along spherical surfaces using spherical harmonics $Y_\ell^m(\theta,\phi)$ where $\ell$ is the spherical harmonic degree, and $m$ is the azimuthal order.  In the radial direction, we expand all variables using  Chebyshev polynomials $T_n(r)$, where $n$ is the degree of the Chebyshev polynomial.  
Time-integration is carried out in the spectral configuration using a hybrid implicit-explicit approach.  A Crank-Nicolson method is used for linear terms, and an Adams-Bashforth approach is employed for the nonlinear terms.  Both components of the time-stepping are 2nd-order in time.  

The numerical algorithm for {\em Rayleigh} is similar to the well-known Anelastic Spherical Harmonic (ASH) code \citep{clune99,brun04}.  However, Rayleigh has been designed to be a bit more flexible (with a Cartesian option) and more scalable (somewhat higher parallel efficiency) than ASH.  It is also somewhat more accessible for both students and the community; the plan is to distribute it publicly by means of the Computational Infrastructure for Geophysics (CIG; see http://geodynamics.org/cig).

Our study is concerned with convection as it manifests deep within stellar interiors, far removed from the photospheric surface where radiative processes may contribute considerably to the energetics of the convection.  In such a region of the star, plasma motions are subsonic and perturbations to thermodynamic variables are small compared to their mean, horizontally averaged values.  Under such conditions, the anelastic approximation, which we employ, provides a convenient means of describing the thermodynamic background state of the system \citep{gough69,gilma81b}.

Under this approximation, thermodynamic variables are linearized about a spherically symmetric reference state with density $\avg{\rho}$, pressure $\avg{P}$, temperature $\avg{T}$, and specific entropy $\avg{S}$.  Here we use an adiabatic reference state so $\del \avg{S} = 0$.  Fluctuations about this state are denoted as $\rho$, $P$, $T$, and $S$.  A further consequence of the anelastic approximation is that the mass flux is solenoidal, reducing the continuity equation to
\begin{equation}  
  \label{eq:continuity}
  \del \cdot(\avg{\rho}\vv) = 0,
\end{equation}
where $\vv$ is the velocity vector expressed in spherical coordinates: $\vv = (v_r,v_{\theta},v_{\phi})$.  The lack of any time derivative in Equation (\ref{eq:continuity}) means that sound waves are naturally filtered out as a consequence of this approach.  The divergence-free constraint for the mass flux is enforced by projecting $\vv$ onto poloidal and toroidal streamfunctions ($W$ and $Z$ respectively), such that
\begin{equation}  
	\avg{\rho}\vec{v} = \del\cross\del\cross(W \rhat)+\del\cross(Z \rhat).
\end{equation}
The unit vector in the radial direction is indicated by $\rhat$.  The momentum equation is given by
\begin{equation}  
  \label{eq:momentum}
  \avg{\rho}\frac{\pd \vv}{\pd t} + \avg{\rho} \left(\vv \bdot \del\right) \vv
  =  -\del P + \rho \ggrav - \del \cdot \scrD,
\end{equation}
where $\ggrav$ is the gravitational acceleration.  The viscous stress tensor $\scrD$ is given by 
\begin{equation}  
\label{eq:stress}
  {\cal D}_{ij} = -2 \avg{\rho} \nu \left[e_{ij}
    - \frac{1}{3}(\del \cdot \vv)\delta_{ij} \right],
\end{equation}
where $e{_{ij}}$ is the strain rate tensor, the kinematic viscosity is denoted by $\nu$, and $\delta_{ij}$ is the Kronecker delta. Our thermal energy equation is given by
\begin{equation} 
  \label{eq:entropytwo}
  \avg{\rho}\avg{T}\frac{\pd S}{\pd t} +   \avg{\rho}\avg{T} ~ \vv \bdot \del S = 
  \del \cdot [\kappa \avg{\rho} \avg{T} \del S] 
                    + 2 \avg{\rho}\nu \left[e_{ij}e_{ij} - \frac{1}{3}(\del \cdot
\vv)^2\right]+Q,
\end{equation}
where the thermal diffusivity denoted by $\kappa$.  Both $\kappa$ and $\nu$ are assumed to be constant, independent of $r$, $\theta$, $\phi$, and $t$.  Sources of internal heating and cooling are encapsulated by $Q$ (see Sec.\ \ref{sec:numex}).  A linearized equation of state closes our set of equations and is given by
\begin{equation} 
  \frac{\rho}{\avg{\rho}} = \frac{P}{\avg{P}} - \frac{T}{\avg{T}}
    =  \frac{P}{\gamma \avg{P}} - \frac{S}{c_p}.
\end{equation}
assuming the ideal gas law
\begin{equation} 
  \avg{P} = \scrR \avg{\rho} \avg{T}.
\end{equation}
The specific heat at constant pressure is denoted by $c_p$, $\scrR$ is the gas constant, and $\gamma$ is the adiabatic index of the gas.  

\subsection{The Numerical Experiments}\label{sec:numex}
We have constructed a series of fourteen stellar convection zone models designed to explore how the convective kinetic energy scales as we decrease the dissipation, approaching solar parameter regimes as discussed in Sec.\ \ref{sec:intro}.  We focus in particular on how this scaling depends on the Prandtl number $P_r$, and on the prescription for cooling at the upper boundary.  These simulations are summarized in Table \ref{simsum}.

Each of our models is constructed using a polytropic background state following \cite{jones11}.  This approach has the advantage that the thermodynamic background may be specified analytically, making it easily reproducible.  All the simulations reported here have the same background state, which is completely specified by seven numbers: the inner radius of the shell $r_i = 5.0 \times 10^{10}$ cm (0.718 $R$), the outer radius of the shell $r_o = 6.586 \times 10^{10}$ cm (0.946$R$), the polytropic index $n_p = 1.5$, the number of density scale heights occurring within the shell $N_\rho = 3$, the mass interior to the shell $M_i = 1.989 \times 10^{33}$ g, the density at the inner boundary $\rho_i = 1.805 \times 10^{-1}$ g cm$^{-3}$, and $c_p = 3.5 \times 10^8$ erg g$^{-1}$ K$^{-1}$.  We note that when constructed using these parameters, the thermodynamic background state closely resembles that of the Sun (FH16).

For all simulations, we have adopted impenetrable and stress free boundary conditions such that
\begin{equation}
\label{eq:vboundary}
\left.v_r=\frac{\partial (v_{\theta}/r)}{\partial r}=\frac{\partial (v_{\phi}/r)}{\partial r}=0\right|_{r = r_i,r_o}.
\end{equation}
For the C series of simulations listed in Table \ref{simsum}, the radial entropy gradient is forced to vanish at the lower boundary of the convection zone, and the entropy perturbations are forced to vanish at the upper boundary:
\begin{equation}
\label{eq:sboundary}
\left.\left.\frac{\partial S}{\partial r} = 0\right|_{r=r_i},~~S = 0\right|_{r=r_o}.
\end{equation}
Thus, there is no diffusive entropy flux across the lower boundary.  Heat is deposited into this system by $Q$ instead, which drops to zero at the upper boundary.  In all simulations, we adopt a functional form for $Q$ that depends only on the background pressure gradient such that
\begin{equation} 
\label{eq:heating}
Q(r) = \alpha (\avg{P}(r)-\avg{P}(r_{o})).
\end{equation}     
The normalization constant $\alpha$ is chosen so that
\begin{equation}
\label{eq:heating2}
L_\odot = 4\pi \int_{r_i}^{r_o} Q(r) r^2 dr, 
\end{equation}
where $L_\odot$ is the solar luminosity.  We then define a corresponding radial energy flux $F_r(r)$ as follows:
\begin{equation}
\label{eq:Fr}
F_r(r) = F_* - \frac{1}{r^2}\int_{r_i}^{r} Q(x) x^2 dx.
\end{equation}
where $F_* = L_\odot / (4\pi r^2)$.   We give this flux the subscript $r$ because we identify it with the flux due to radiative diffusion.  It is defined such that $Q = -\dv (F_r \rhat)$, $F_r(r_i) = F_*$ and $F_r(r_o) = 0$.  Thus, radiative diffusion carries the entire solar luminosity in to the computational domain through the bottom boundary and no flux through the top boundary.  This implies a convergence of $F_r$ that establishes and sustains the convection.

\begin{figure}
\begin{center}
\leftline{\includegraphics[width=0.5\columnwidth]{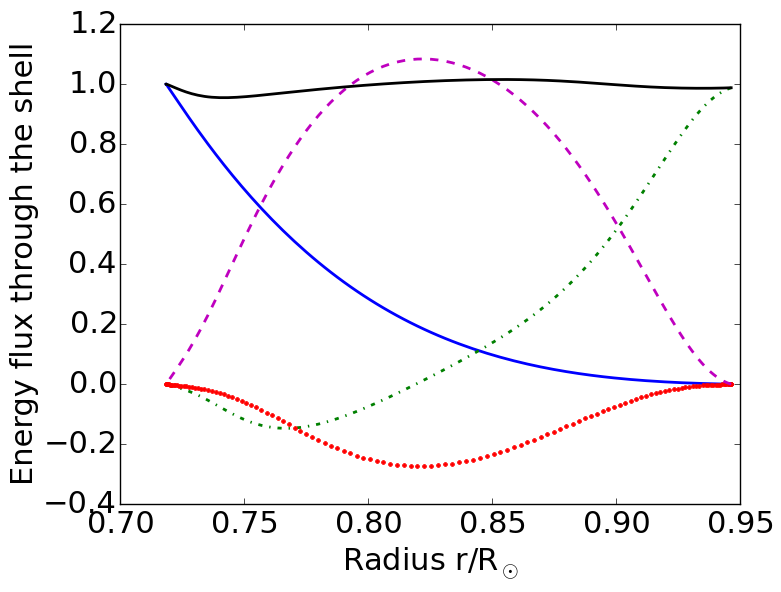}}
\vskip-0.37\columnwidth
\rightline{\includegraphics[width=0.5\columnwidth]{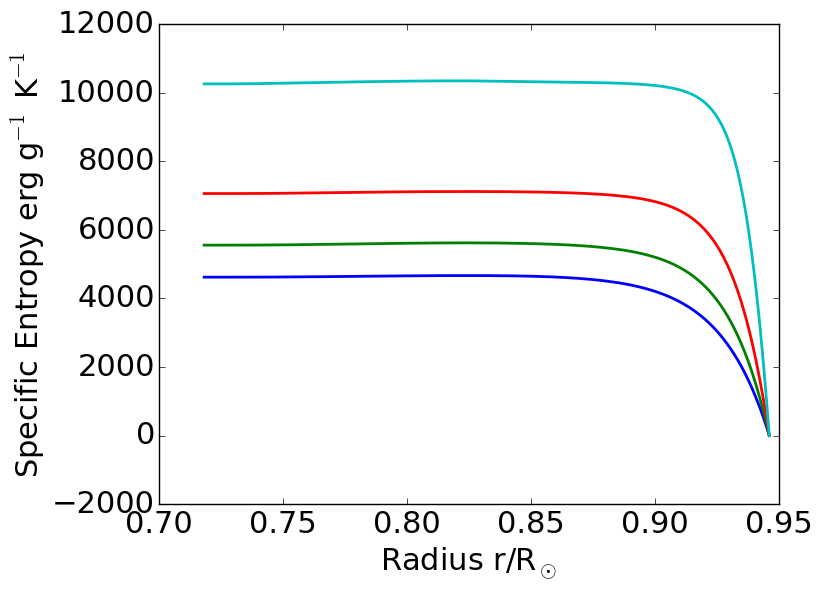}}
\end{center}
\caption{(Left) Flux balance for case C1. The y axis represents the energy flux normalized by the solar flux $F_*$ and the x axis represents the radius normalized by the solar radius $R$. The red dotted line represents the kinetic energy flux $F_k$, the green dash-dotted line is the conductive flux $F_c$, the purple dashed line is the enthalpy flux $F_e$, the blue solid line is the radiative flux $F_r$, and the black solid line near unity represents the sum of all the fluxes. The graph shows a rise in the enthalpy flux in the middle of the convection zone.  (Right) This graph shows the specific entropy $S$ over the convection zone for different values of $\kappa$ while $\nu$ is held constant at $8 \times 10^{12}$ cm$^2$ s$^{-1}$. The blue line represents $\kappa = 8 \times 10^{12}$ cm$^2$ s$^{-1}$ (case C1), the green line represents $\kappa=6 \times 10^{12}$ cm$^2$ s$^{-1}$ (Case C2), the red represents $\kappa = 4 \times 10^{12}$ cm$^2$ s$^{-1}$(Case C3), and the tiel represents $\kappa = 2 \times 10^{12}$ cm$^2$ s$^{-1}$ (Case C4). As $\kappa$ decreases the graph shows a systematic increase in $\Delta S$.\label{fig:Fluxbalance}}
\end{figure}

To illustrate how this occurs, we consider the other components of the radial heat flux, which include the convective enthalpy flux $F_e$, the kinetic energy flux $F_k$, and the conductive flux $F_c$, defined as follows:
\begin{eqnarray}
F_e & = \avg{\rho} C_p \left< v_r T \right> \label{eq:Fe} \\
F_k & = \frac{1}{2} \avg{\rho} \left< v_r \left|v\right|^2\right> \\
F_c & = \kappa \avg{\rho} \frac{\pd \left<S\right>}{\pd r} \label{eq:Fc} ~~~.
\end{eqnarray}
Angular brackets denote averages over horizontal surfaces ($\theta$ and $\phi$).   In a steady state, these fluxes must balance such that $F_e + F_k + F_r + F_c = F_*$.

Each simulation is initialized using a small random thermal fluctuation.  As time proceeds, the convergence of the radiative heat flux heats the convection zone (CZ), raising the adiabat and establishing a negative $\pd S/\pd r$ near the surface, as illustrated in the right panel of Fig.\ \ref{fig:Fluxbalance}.  This superadibatic entropy gradient excites convective motions that grow to dominate the heat transport through the mid CZ by means of the resolved convective fluxes $F_e$ and $F_k$.  The resulting balance is shown in Fig.\ \ref{fig:Fluxbalance} for Case C1 as an example.  Each simulation was evolved until the kinetic energy reached a statistically steady  (equilibrated) state, and then further evolved for at least one thermal diffusion time (some were evolved up to ten thermal diffusion times). 

For the C series of simulations, heat exits the upper boundary via the conductive flux, $F_c$.  There, the steepness of the equilibrated entropy gradient is entirely dependent upon the thermodynamic background state, the value of the thermal diffusivity $\kappa$, and the chosen luminosity. Specifically,
\begin{equation}
\label{eq:sgradient}
\left.\frac{\partial S}{\partial r}\right|_{r=r_o} = \frac{L_\odot}{4\pi r_o^2\kappa \avg{\rho}(r_o) \avg{T}(r_o)}  ~~~.
\end{equation}
This leads to a variation among simulations in the entropy difference across the CZ, $\Delta S = \left<S\right>_{r=r_i} - \left<S\right>_{r=r_o}$.  Equilibrated values of $\Delta S$ for all simulations are included in Table \ref{simsum}.

For the G series of simulations, eq.\ (\ref{eq:sgradient}) no longer holds.  There we have imposed another flux component $F_g$ that contributes to the heat transport through the outer surface (see Table \ref{simsum}).  This is intended to mimic heat transport by small-scale surface convection as discussed in \S\ref{sec:fg}.  We have also modified the upper thermal boundary condition in the G series, fixing the entropy gradient, $\pd S/\pd r$, as opposed to the entropy $S$, as expressed in eq.\ (\ref{eq:sboundary}).  Apart from these differences, the $G$ series was initiated and evolved in a similar manner as the $C$ series.  The difference between the different members of the C and G series are the values of $\kappa$ and $\nu$, as indicated in Table \ref{simsum}.

\begin{table}
\caption{Simulation Summary.\label{simsum}}

\begin{center}
\begin{tabular}{lccccc}
\hline
Case & $\kappa$ & $\nu$ & $P_r$ & Surface & $\Delta S$ \\
     & ($10^{12}$ cm$^2$ s$^{-1}$) & ($10^{12}$ cm$^2$ s$^{-1}$) & & Flux & erg g$^{-1}$ K$^{-1}$ \\
\hline
C1 & 8 & 8 & 1 & $F_c$ & 4620 \\ 
C2 & 6 & 8 & 1.33 & $F_c$ & 5560 \\ 
C3 & 4 & 8 & 2 & $F_c$ & 7060 \\ 
C4 & 2 & 8 & 4 & $F_c$ & 10300 \\ 
C5 & 6 & 6 & 1 & $F_c$ & 5450 \\ 
C6 & 4 & 4 & 1 & $F_c$ & 6780 \\ 
C7 & 2 & 2 & 1 & $F_c$ & 8870 \\ 
\hline
G1 & 8 & 8 & 1 & $F_c + F_g$ & 3040 \\
G2 & 6 & 8 & 1.33 & $F_c + F_g$ & 2690 \\
G3 & 4 & 8 & 2 & $F_c + F_g$ & 2422 \\
G4 & 2 & 8 & 4 & $F_c + F_g$ & 2033 \\
G5 & 6 & 6 & 1 & $F_c + F_g$ & 2704 \\
G6 & 4 & 4 & 1 & $F_c + F_g$ & 2312 \\
G7 & 2 & 2 & 1 & $F_c + F_g$ & 1900 \\
\hline
\end{tabular}
\end{center}
\end{table}

\section{Results: The Role of the Prandtl Number}\label{sec:Prandtl}

The Prandtl Number ($P_r$) is defined as $P_r=\nu/\kappa$ (Sec.\ \ref{sec:model}). FH16 investigated the scaling of the kinetic energy with $\kappa$ in global, non-rotating convection simulations for the special case in which the value of $P_r$ is fixed at unity.  They found that the characteristic velocity amplitude $U$ first increased slightly as $\kappa$ was decreased but then asymptoted toward a constant value in the limit $\kappa \rightarrow 0$.  Here we consider the case in which $\nu$ is held fixed as $\kappa$ is decreased (implying an increasing value of $P_r$) and we find that in this case $U$ systematically decreases.

In figure \ref{fig:Fluxbalance} (left) we show the flux balance across the convection zone for an illustrative simulation in the C series, case C1. The figure shows that $F_r$ is the main transporter of the solar luminosity at low values of $r$ (see Sec.\ \ref{sec:numex}). Near the center of the convection zone ($r \approx 0.83 R$) $F_e$ transports nearly all the solar luminosity to the sun's surface, while at the boundaries ($r=0.70$ and $r=0.95$) $F_e$ transports none of the solar luminosity. $F_c$ takes over as the main carrier of the solar luminosity as $r \rightarrow r_o$.

\begin{figure}
\begin{center}
\leftline{\includegraphics[width=0.5\columnwidth]{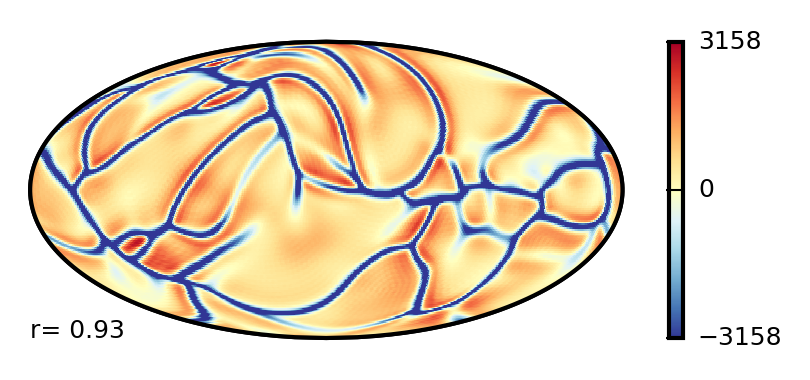}}
\vskip-0.235\columnwidth
\rightline{\includegraphics[width=0.5\columnwidth]{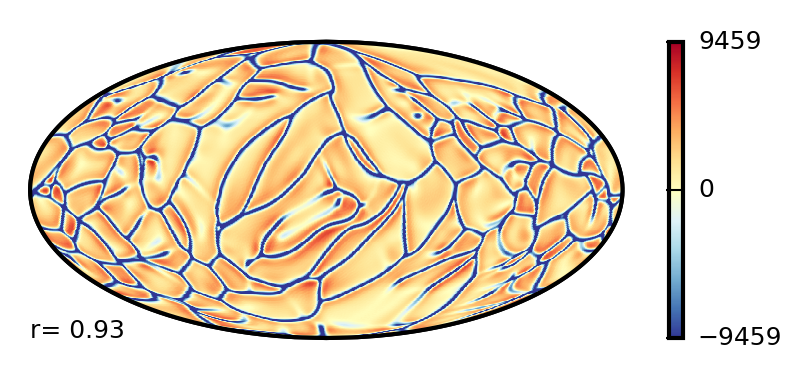}}
\end{center}
\caption{Radial velocity snapshots at $r = 0.93 R$ for (Left) Case C1 and (Right) Case C4 (Molleweide projection).  The red indicates movement up and out of the sun while the blue indicates movement down and into the sun, as indicated by the color bars (saturation levels in cm s$^{-1}$).\label{fig:moll1}}
\end{figure}

In the right panel of figure 1 we look at the specific entropy over $r$ for various $\kappa$ values. As the value of $\kappa$ decreases (blue being case C1 and tiel being case C4) we see an increase in the change of the specific entropy from the bottom to the top of the convection zone. As noted in Sec.\ \ref{sec:numex}, this change occurs because a lower $\kappa$ value causes an increase in the entropy gradient between the center and top of the convection zone (see eq.\ \ref{eq:sgradient}). 

\begin{figure}
\begin{center}
\leftline{\includegraphics*[width=0.5\columnwidth]{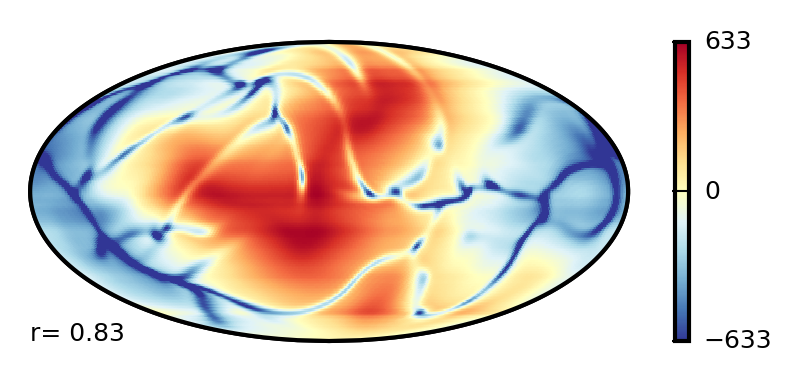}}
\vskip-0.24\columnwidth
\rightline{\includegraphics[width=0.5\columnwidth]{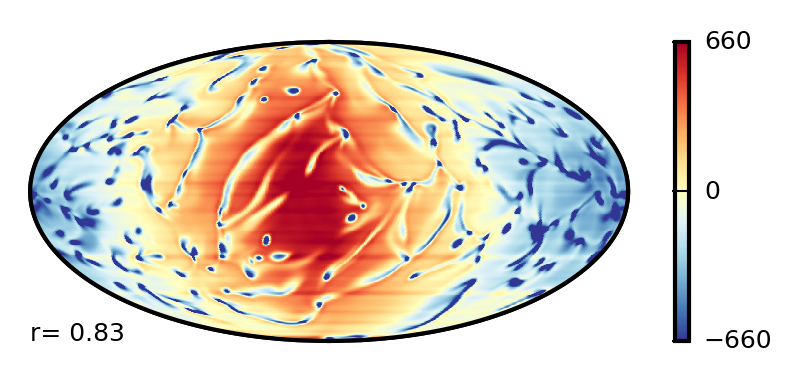}}
\end{center}
\caption{Specific entropy snapshots at $r = 0.83$ for (Left) Case C1 and (Right) Case C4 (Molleweide projection).  Red and blue colors denote higher and lower entropy, as indicated by the color bars (saturation values given in erg g$^{-1}$ K$^{-1}$).\label{fig:moll2}}
\end{figure}

Figures \ref{fig:moll1} and \ref{fig:moll2} compare the structure of the convection in two simulations with the same value of $\nu$ but different values of $\kappa$ (Cases C1 and C4).  A lower $\kappa$ value increases both the Rayleigh number and the Prandtl number, causing the convection to become more intricate and apparently more turbulent (Fig.\ \ref{fig:moll1}).  Both the characteristic size of the convection cells and the characteristic width of downflow lanes is smaller for Case C4 (right panel) than for Case C1 (left panel).  The latter can be attributed to the thinner thermal boundary layer (see the right panel of Fig. \ref{fig:Fluxbalance} and Sec.\ \ref{sec:discussion}).

Similar results are observed in the snapshots of the specific entropy in the mid CZ for cases C1 and C4 (Fig.\ \ref{fig:moll2}).   The smaller value of $\kappa$ yields a more intricate, smaller-scale structure with larger thermal fluctuations.   In the right panel of Fig.\ \ref{fig:moll2} smaller and sharper features can be seen than in the left panel. This is particularly noticeable for the cold spots (blue) which represent small, cool plumes of fluid that sink down from the surface layers.

Also notable in Fig.\ \ref{fig:moll2} is a pronounced dipole convective mode, visible here as a pattern with longitudinal wavenumber $m=1$.  This is a common feature for convection in non-rotating spherical geometries, related to the linearly preferred convective modes \citep{chand61}.  It is analogous to the box mode or ``wind'' that commonly arises in studies of convection in Cartesian geometry.  Its presence is further promoted by the fixed-heat flux lower boundary condition, which favors convective modes with long horizontal wavelengths \citep{chapm80,depas81}. The dipolar mode is most visible in temperature or entropy plots but it can be seen in radial velocity plots as well (e.g.\ Fig.\ \ref{fig:moll1}). The primary effect of the dipole mode is to sweep small-scale, plume-like structures out of the dipolar upflow region, thereby causing them to cluster in the region of convergence associated with the dipolar downflow region.   This effect is not observed in rotationally constrained convection because the presence of rotation tends to favor convective modes with higher azimuthal wavenumber (high-$m$ banana cells).

However, due to the spherical symmetry of the equations, the orientation of this dipole mode should be random; there is no particular reason for it to be nearly perpendicular to the rotation axis as it is in both cases highlighted in Fig.\ \ref{fig:moll2} (see also Fig.\ \ref{fig:moll3} below).  This may be a coincidence; other simulations (not shown here) do exhibit a wide range of orientations for the dipole mode.  However, it's also possible that the numerical grid introduces a slight symmetry breaking that is enough to influence the orientation of the dipole mode.  In any case, we do not expect the orientation of the dipole mode to significantly influence our results.  The faint horizontal striping pattern in Fig.\ \ref{fig:moll2} appears to be an artifact arising from the python plotting program; it is not present in the raw data (before the Molleweide projection).

If $\nu$ is decreased along with $\kappa$, this also leads to a flow that appears more turbulent.  This can be seen by comparing the left panel of Fig.\ \ref{fig:moll3} to Fig.\ \ref{fig:moll1} and the right panel of Fig.\ \ref{fig:moll3} to Fig.\ \ref{fig:moll2}.   This greater level of turbulence is usually attributed to an increase of the Rayleigh number, which scales as $(\nu \kappa)^{-1}$ (FH16).  

The apparent increase in the level of turbulence for cases with lower dissipation also suggests a greater convective amplitude, $U$. However, in Fig.\ \ref{fig:vrms} we see that this is not necessarily the case.  In the left panel we have plotted the root mean square (rms) value of the convective velocity versus 1/$\kappa$ for the C series of simulations.  We use this rms velocity as a measure of $U$.  The results for fixed $P_r$ agree with those presented by FH16 and show $U$ increasing toward a constant value as $\kappa$ is decreased.  However, the results for fixed $\nu$ show a decrease in $U$ with decreasing $\kappa$.  

The level of turbulence is typically quantified by the Reynolds number, $R_e = U L / \nu$, where $L$ is a characteristic length scale of the flow.  If we take $L$ to be the depth of the layer, then a decrease of $U$ at fixed $\nu$ also implies a decrease in $R_e$.  This conclusion also holds if we use a more meaningful measure of the turbulent length scale $L$, such as the integral scale of the flow or the Taylor microscale.   Either of these measures would imply a decrease in $L$ with decreasing $\kappa$, further reducing $R_e$.  This can be verified by considering the power spectra shown in the right panel of Fig.\ \ref{fig:vrms}.  As $\kappa$ decreases the power spectrum shifts toward higher values of the wave number ($\ell$). This indicates smaller eddies, which in turn implies a decrease in $L$.

Thus, we arrive at the somewhat surprising conclusion that the flow at low values of $\kappa$ and fixed $\nu$ is actually less turbulent (in the sense of smaller $R_e$), despite its more intricate, smaller-scale structure, and higher Rayleigh number.

\begin{figure}
\begin{center}
\leftline{\includegraphics*[width=0.5\columnwidth]{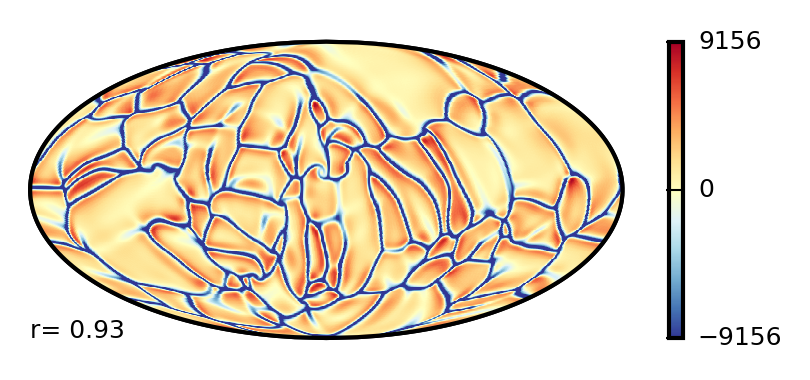}}
\vskip-0.245\columnwidth
\rightline{\includegraphics*[width=0.5\columnwidth]{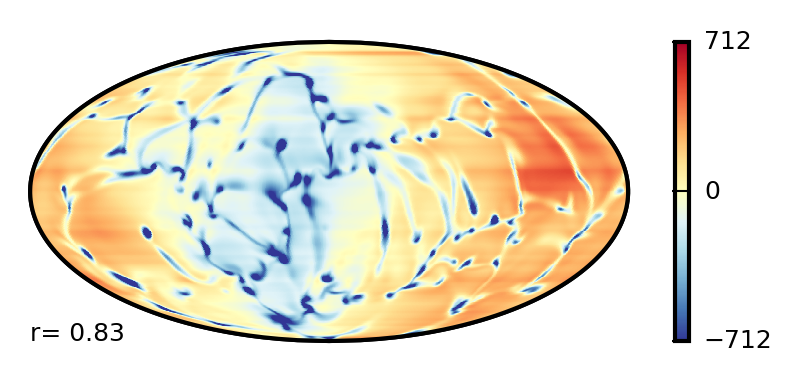}}
\end{center}
\caption{Snapshots of (left) the radial velocity at $r = 0.93R$ and (right) the specific entropy at $r = 0.83R$ for Case C6 ($\kappa = \nu = 4 \times 10^{12}$ cm$^2$ s$^{-1}$).  Color bars are as in Figs.\ \ref{fig:moll1} and \ref{fig:moll2}.\label{fig:moll3}} 
\end{figure}

\begin{figure}
\begin{center}
\leftline{\includegraphics*[width=0.5\linewidth]{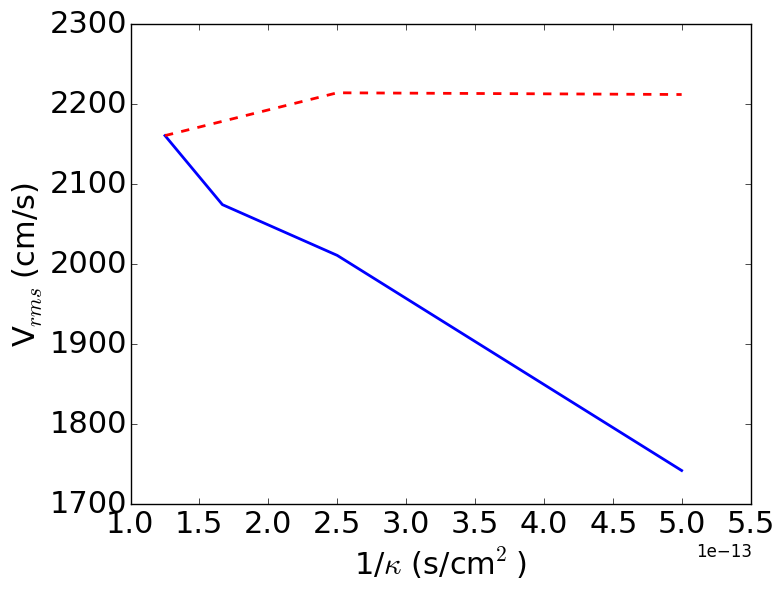}}
\vskip-0.385\columnwidth
\rightline{\includegraphics*[width=0.49\columnwidth]{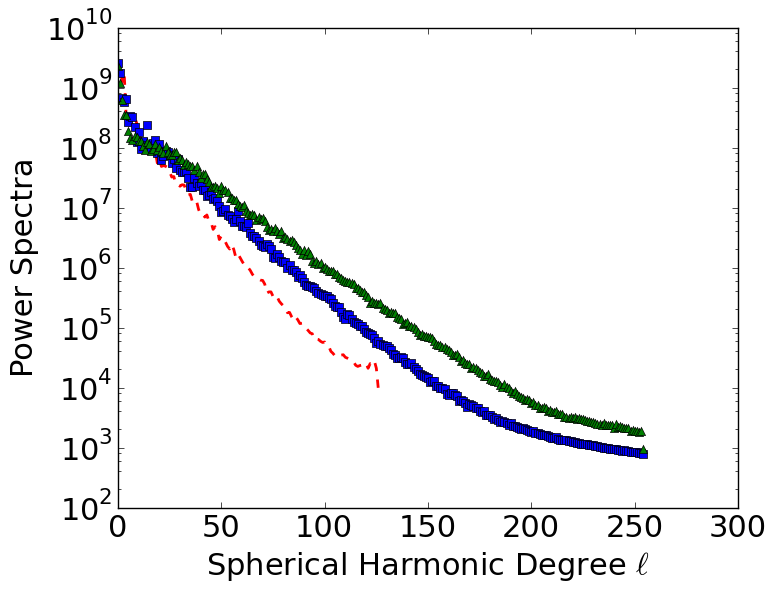}}
\end{center}
\caption{(Left) rms velocity for different values of $\kappa$ in Cases C1 through C7. The 
solid blue line shows the rms velocity versus $1/\kappa$ for fixed $\nu$ (C1, C2, C3, C4) and
the dashed red line shows the rms velocity versus $1/\kappa$ for fixed $P_r$ (C1, C5, C6, C7).
(Right) The velocity power spectrum (units cm$^2$ s$^{-2}$) at $r = 0.93 R$ for Cases 
(red dashed) C1, (blue boxes) C3, and (green triangles) C4 ($\kappa = 8$, 4, and 
2 $\times 10^{12}$ cm$^2$ s$^{-1}$ respectively), 
versus spherical harmonic degree.\label{fig:vrms}}
\end{figure}

\section{Heat Transport by Small-Scale Surface Convection}\label{sec:fg}

Since the entropy fluctuations responsible for the buoyant driving of convection originate mainly in the upper thermal boundary layer, one might expect convective velocity amplitudes to be sensitive to the details of how this region is modeled (see \S\ref{sec:discussion}).  In the Sun, heat transport in the outer 10 Mm of the convection zone ($r > 0.98$) is dominated by relatively small-scale convective motions ranging in size from about 1-2 Mm (granulation) to about 30-35 Mm (supergranulation).  Near the solar photosphere, convective transport transitions to radiative heat transport that carries energy away from the solar surface and into interplanetary space.  No existing global solar convection simulation has the spatial resolution, temporal resolution, and physical ingredients (radiative transfer, non-ideal equation of state, compressibility) to realistically capture the complex dynamics in this outermost region of the Sun.  So, we must rely on modeling.

The approach we have taken for the simulations presented in \S\ref{sec:Prandtl} parallels that taken by many previous studies, dating back to the pioneering work of \cite{gilma83} and \cite{glatz85a}; see \cite{miesc05} and \cite{jones11} for further references and discussion.  In particular, we modeled the heat flux by unresolved, small-scale convection in the near surface layers as a turbulent thermal diffusion, $F_c$, that operates on the entropy gradient as opposed to the temperature gradient (see eqs.\ \ref{eq:entropytwo} and \ref{eq:Fc}).  This mimics the tendency for efficient convection to mix entropy, establishing a nearly adiabatic stratification.  

Though physically justified, the representation of subgrid-scale (SGS) heat transport as a thermal conduction is an approximation that needs to be evaluated.  One consequence of this model is that it ties the SGS heat flux to the background entropy gradient, $\pd S/\pd r$, which has important implications for the overall dynamics of the convection zone (see \S\ref{sec:discussion}).

\begin{figure}
\begin{center}
\leftline{\includegraphics[width=0.5\columnwidth]{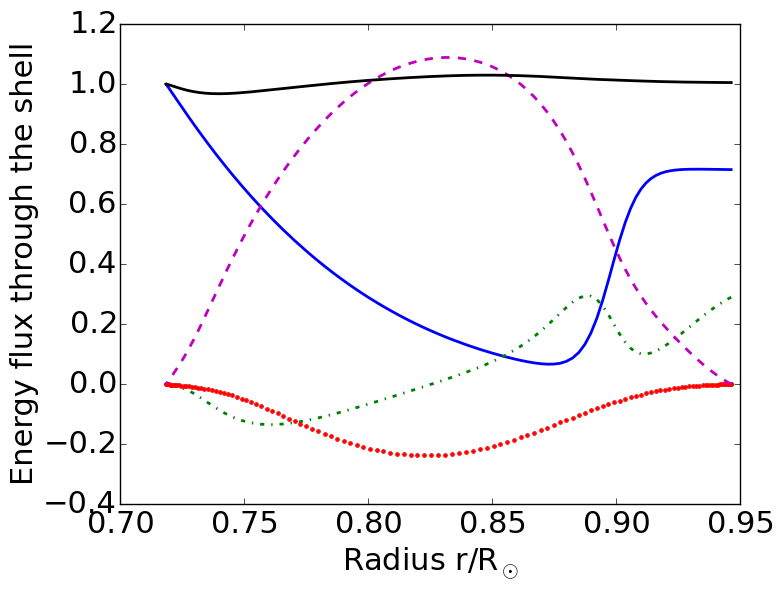}}
\vskip-0.38\columnwidth
\rightline{\includegraphics[width=0.5\columnwidth]{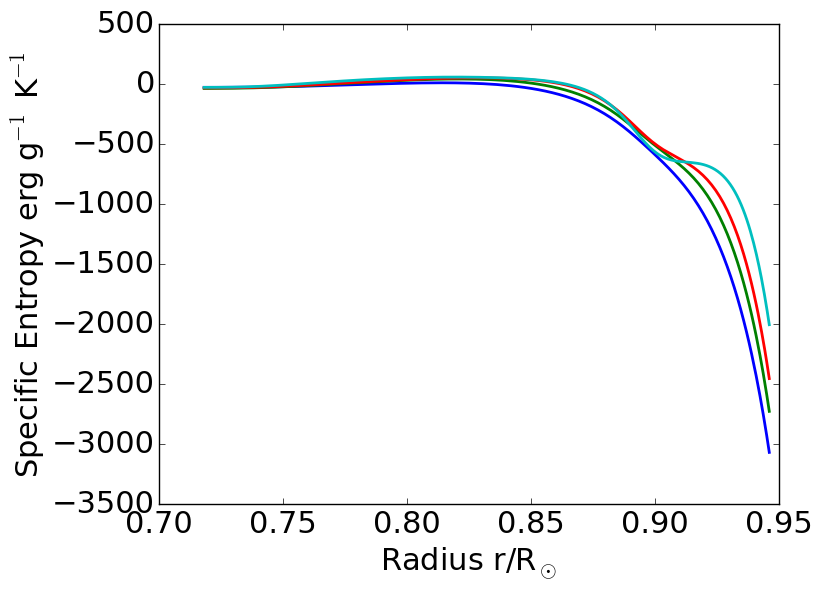}}
\end{center}
\caption{(Left) Similar to the left panel of Fig.\ \ref{fig:Fluxbalance}, but for Case G3.  However, here the contribution from $F_g$ has been combined with the radiative flux $F_r$, so the solid blue curve represents $F_r + F_g$.  (Right) Similar to the right panel of Fig.\ \ref{fig:Fluxbalance} but for cases (blue) G1, (green) G2, (red) G3, and (tiel) G4.  Here the magnitude of $S$ at the top boundary decreases systematically with decreasing $\kappa$.\label{fig:fbal_fg}}
\end{figure}

In order to assess these implications further, we have initiated another series of simulations with a different model for the SGS heat flux in the surface layers.  This model consists of an imposed radial heat flux, $F_g$, that is independent of the background stratification and the convective flow field.  Its functional form is given by
\begin{equation}
F_g = \frac{a}{2} \left(\frac{L}{4\pi r^2}\right) \left(1 + \tanh\left[\frac{r - r_g}{d_g}\right]\right)  ~~~.
\end{equation}
If the parameter $a$ is set to unity, then this formulation stipulates that $F_g$ will carry the entire solar luminosity $L$ in the region $r_g \lesssim r \leq r_o$.  For $r \ll r_g$, $F_g \rightarrow 0$.  The width of the transition at $r = r_g$ is governed by the parameter $d_g$.  Here we use $r_g = 0.90R$ and $d_g = (r_o-r_i)/30$.

For this series of simulations, which we refer to as the G series ($F_g \neq 0$), we also use a different upper thermal boundary condition.  Instead of fixing $S$ as in the $C$ series, we fix the entropy gradient $\pd S/\pd r$ at a value of $-2 \times 10^{-6}$ cm K$^{-1}$ s$^{-2}$.  Since we do not zero out $\kappa$, this implies that we we still have a nonzero thermal conduction, $F_c$, that contributes to the heat flux through the top boundary.  However, since $\pd S/\pd r$ is fixed, $F_c(r=r_o) \rightarrow 0$ as $\kappa \rightarrow 0$.  

This brings us to the significance of the parameter $a$.  This is an amplitude factor that we adjust to ensure that the total flux through the outer boundary ($r = r_o$) is equal to the solar flux; $F_g(r=r_o) + F_c(r=r_o) = L_\odot / (4 \pi r_o^2)$.  As $\kappa \rightarrow 0$, more of the flux through the boundary must be carried by $F_g$, so $a \rightarrow 1$.  However, for the values of $\kappa$ considered here, the contribution from $F_c$ is still significant, implying $0  < a < 1$.

This is illustrated in the left panel of Fig.\ \ref{fig:fbal_fg}, which shows the flux balance for Case G3.  Here, $F_c$ transports about 28\% of the solar luminosity through the surface, leaving about 72\% to be carried by $F_g$ ($a \approx 0.72$).  The right panel shows the resulting entropy stratification.  Here the spread in $\Delta S$ is smaller than in the C series and shows an opposite trend, decreasing slightly with decreasing $\kappa$ instead of increasing.  Furthermore, unlike the C series, the entropy profile in the G series will become independent of $\kappa$ in the limit $\kappa \rightarrow 0$.

\begin{figure}
\begin{center}
\leftline{\includegraphics*[width=0.5\columnwidth]{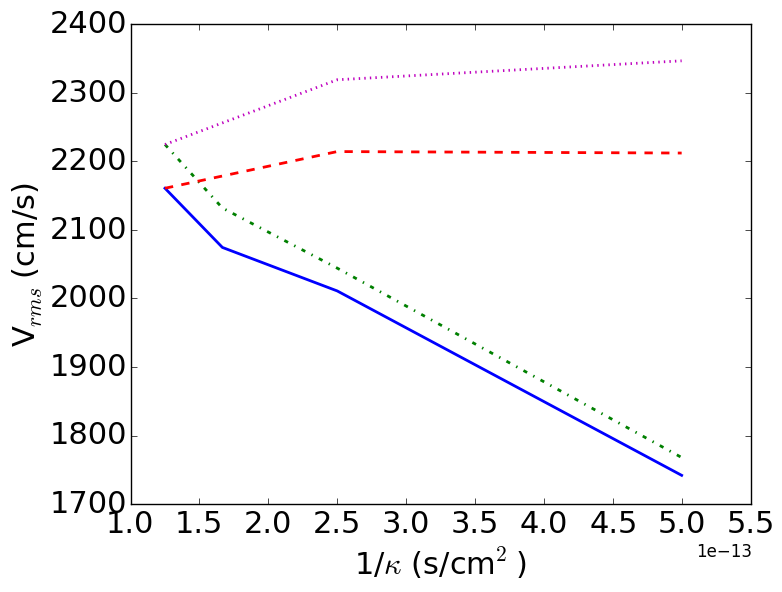}}
\vskip-0.385\columnwidth
\rightline{\includegraphics*[width=0.5\columnwidth]{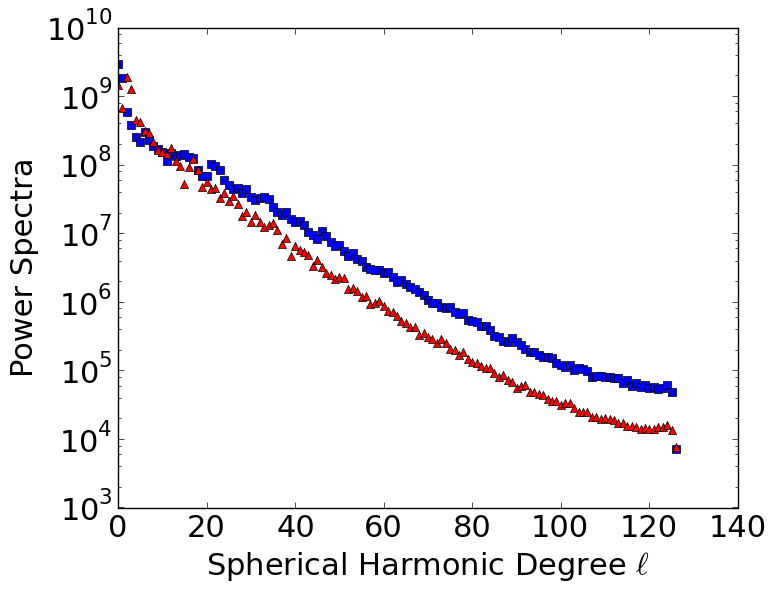}}
\end{center}
\caption{(Left) As in Fig. \ref{fig:vrms} but including results from all fourteen simulations (C series and G series), with and without the granulation flux $F_g$.  Solid blue and dashed red lines are as in Fig.\ \ref{fig:vrms}.  Their counterparts for nonzero $F_g$ are indicated by the dash-dotted green (G1, G2, G3, G4) and dotted purple (G1, G5, G6, G7) lines.  (Right) Velocity power spectrum at $r = 0.93 R$ for Cases C2 (blue squares) and G2 (red triangles).  The units are cm$^2$ s$^{-2}$.\label{fig:vrms_fg}}
\end{figure} 

Remarkably, even with this very different treatment of the upper boundary layer, the results are similar.  This is demonstrated for the rms velocity amplitude in the left panel of Fig.\ \ref{fig:vrms_fg}.  Though the velocity amplitudes are systematically higher in the G series of simulations, the trends are similar to the C series.  In particular, $U$ decreases with decreasing $\kappa$ if $\nu$ is held fixed but increases and asympotes toward a constant value if $P_r$ is held fixed.   The velocity spectra are also similar, though somewhat steeper in the G series (Fig.\ \ref{fig:vrms_fg}, right panel).  This can likely be attributed to the greater width of the thermal boundary layer for nonzero $F_g$, which is expected to produce wider plumes (\S\ref{sec:discussion}).

Thus, we conclude that the results, and in particular the velocity amplitudes, are insensitive to the details of how the upper thermal boundary layer is modeled.  This is consistent with the high-resolution, global, non-rotating simulations of \cite{hotta14a}.  They compared several simulations with different surface cooling and different locations for the upper boundary, $r_o = 0.96 R$ and $r_o = 0.99R$.  The latter employed an unprecedented spatial resolution in order to explicitly capture small-scale convective motions near the surface (though even this simulation did not have enough resolution to extend the convective spectrum all the way down to granulation).  They found that various measures of the velocity field in the mid convection zone, including the velocity amplitude and spectrum,  were similar in the different cases considered and thus insensitive to the detailed structure and dynamics of the upper boundary layer. 

\section{Discussion}\label{sec:discussion}

When interpreting our results, we must keep in mind the target parameter regime.  As discussed in \S\ref{sec:fg}, the conductive flux $F_c$ is a model that is intended to represent SGS convection in the solar surface layers.  As such, there is no value of $\kappa$ that corresponds to the microphysical (plasma) thermal diffusion in the actual Sun.  However, there is a thermal diffusion due to radiative heat transport (modeled here implicitly by $Q$), which is proportional to $\kappa_r \del T$.  Thus, we may loosely take the value of $\kappa_r \sim 10^7$ cm$^2$ s$^{-1}$ as our target value for the effective $\kappa$ in the Sun.  Since the values of $\kappa$ used here are on the order of $10^{12}$-$10^{13}$ cm$^2$ s$^{-1}$, we would need to decrease $\kappa$ by about 5-6 orders of magnitude to approach the actual thermal diffusion of the Sun.  The molecular viscosity, $\nu$, in the Sun is even smaller, on the order of 1 cm$^2$ s$^{-1}$ \citep{miesc05}.  So, in order to achieve the true microphysical plasma viscosity of the Sun, we would have to decrease $\nu$ by about 12-13 orders of magnitude.  So, based on the microphysics of the solar plasma, $P_r \ll 1$.  However, as we will argue in \S\ref{sec:summary}, the effective viscosity of the solar plasma may be much higher, potentially placing the Sun in a high-$P_r$ regime.  

This justifies our two paths in parameter space.  We seek to understand the behavior of the system as we decrease $\kappa$ to approach more solar-like parameter regimes.  However, we find that this behavior depends on whether or not we simultaneously decrease $\nu$ as well.

Within this context, the aim of this section is to understand and interpret the principle result of Section \ref{sec:Prandtl}, namely that:
\begin{itemize}
\item R1: $U$ {\em \bf decreases} with decreasing $\kappa$ if $\nu$ is held fixed.
\end{itemize}
where $U$ is the characteristic velocity amplitude of the convection.  If this result holds up then it bodes well for global solar convection simulations, that employ an artificially large value of $\kappa$ and that appear to be overestimating $U$ (\S\ref{sec:intro}).

While we are addressing this principle result, we will also address the other results discussed in sections \ref{sec:Prandtl} and \ref{sec:fg}, which may be summarized as follows:
\begin{itemize}
\item R2: $U$ initially {\em \bf increases} with decreasing $\kappa$ if $P_r$ is held fixed, asymptotically approaching a {\em \bf constant value} as $\kappa \rightarrow 0$.  This result was previously reported by FH16.
\item R3: R1 and R2 still apply if the conductive heat flux through the top boundary is (partially) replaced by a fixed heat flux designed to mimic small-scale surface convection. 
\end{itemize}

Note that the range of $\kappa$ spanned by the simulations presented here is insufficient to provide a robust characterization of the trend found in R1.  In particular, the dependence appears roughly linear (Fig.\ \ref{fig:vrms_fg}), but we cannot rule out other power law exponents or functional relationships.  Similar caveats apply to R3.  By contrast, R2 has indeed been thoroughly and robustly characterized by FH16. 

\subsection{Rayleigh-B\'enard (RB) Convection}\label{sec:RB}

To provide context for understanding results R1 and R2 it is instructive to consider the most well-studied manifestation of thermal convection in the literature, namely Rayleigh-B\'enard (RB) convection.  The physical system consists of a Boussinesq fluid (no density stratification) between two semi-infinite horizontal plates at fixed temperature \citep{chand61,gasti15}.   The mathematical system can be completely specified by two nondimensional numbers, namely the Rayleigh number $R_a$ and the Prandtl number, $P_r$.  In the Boussinesq limit, $R_a = \alpha g D^3 \Delta / (\kappa \nu)$, where $\alpha$ is the thermal expansion coefficient, $g$ is the gravitational acceleration, $D$ is the distance between the plates, and $\Delta$ is the imposed temperature difference.  

The characteristic convective velocity scale, $U$, is typically expressed nondimensionally as the Reynolds number $R_e = U D / \nu$.  Whereas $R_a$ and $P_r$ are system parameters, $R_e$ is a diagnostic output of the numerical or laboratory experiment and, for $R_a >>1$, it is found to scale as 
\begin{equation}\label{eq:sc1}
R_e \propto R_a^a ~ P_r^b
\end{equation}
If the thermal boundary layer is thinner than the viscous boundary layer ($P_r \gtrsim 0.1$) and if it is assumed that there is a local force balance between buoyancy and advection in each convective plume, then the basic mixing length theory suggests $a = 4/9$ and $b = -2/3$ \citep{siggi94}.  However, more sophisticated theories and laboratory experiments paint a more complex picture, with multiple parameter regimes determined by the spatial distribution of the thermal and viscous dissipation and the relative widths of the corresponding boundary layers \citep{ahler09,gasti15}.  Still, in all of these regimes, $a > 0$ and $b < 0$, implying that $R_e$ increases with increasing $R_a$ and decreases with increasing $P_r$.  

This qualitative scaling is straightforward to interpret.  For fixed $P_r$, increasing $R_a$ by, for example, increasing the temperature difference or decreasing the diffusion, will cause the flow to become more turbulent (larger $R_e$).  Meanwhile, increasing $P_r$ while keeping $R_a$ fixed implies that the flow becomes more viscous (larger $\nu$) and laminar (smaller $R_e$).  However, the dimensional scaling is more subtle.  Dimensionally, eq.\ (\ref{eq:sc1}) implies
\begin{equation}\label{eq:sc2}
U \propto \nu^{1-a+b} ~ \kappa^{-(a+b)} = P_r^{1-a+b} ~ \kappa^{1-2a} ~~~~.
\end{equation}
So, for fixed $\nu$, $U$ decreases with decreasing $\kappa$ (R1) only if $a+b < 0$.  This is indeed the case for the mixing length estimate quoted above and for most of the parameter regimes described by \cite{ahler09}.  Meanwhile, $U$ becomes independent of $\kappa$ at fixed $P_r$ (R2) if $a = 1/2$.  The above estimate of $a = 4/9$ is less than $1/2$ so in this scaling regime, $U$ would decrease with decreasing $\kappa$, in contrast with result R2.  However, for $P_r \sim 1-10$ and $R_a \sim 10^5$-$10^7$, the more comprehensive analysis by \cite{ahler09} suggests values of $a$ that are indeed equal to $1/2$.  So the scaling laws quoted in the literature for RB convection are at least qualitatively consistent with our results R1 and R2.

However, a more careful consideration of the underlying physics suggests that the trends we are seeing (relevant for solar convection) are fundamentally different that the trends previously found for RB convection.  In RB convection, the decrease of $U$ with decreasing $\kappa$ for fixed $\nu$ (R1) is accompanied by a decrease in the heat flux through the layer, as quantified by the Nusselt number $N_u$ \citep{siggi94,ahler09}.  Conversely, an increase in $U$ is generally accompanied by an increase in the heat flux.  By contrast, the heat flux in our simulations, and in the Sun, is fixed by the solar luminosity.  We will address this fundamental difference and its implications in \S\ref{sec:lumcon}.

In the RB case, the decrease of $U$ for decreasing $\kappa$ at fixed $\nu$ (R1) can be attributed to what we concisely refer to hereafter as {\em \bf a shift in the power spectrum}.  In RB convection, as in the simulations presented here, the characteristic width of convective plumes is roughly determined by the thickness of the thermal boundary layer which scales approximately as $\kappa^{1/2}$ \citep[][FH16]{ahler09}.  As $\kappa$ decreases, the thermal boundary layer becomes thinner and the plumes become narrower, shifting the entire power spectrum toward higher wavenumber.  At fixed $\nu$, this enhances the viscous dissipation.  It also enhances the turbulent entrainment of surrounding fluid, suppressing the buoyant driving.  The result is a decrease in both $U$ (R1) and the heat flux, $N_u$.  

However, if $\nu$ is decreased in pace with $\kappa$ (fixed $P_r$), then the viscous dissipation can be held in check.  Now one might expect $U$ to be determined by the potential energy available in the upper boundary layer; see FH16 and \S\ref{sec:ass} below.  Note that the shift in the power spectrum and in the buoyancy work toward higher wavenumbers as $\kappa$ is decreased has been demonstrated by FH16.  However, since this is achieved by means of a fixed-$P_r$ path through parameter space, this shift is not associated with a significant increase in the viscous dissipation.

\subsection{The Luminosity Constraint}\label{sec:lumcon}

In section \ref{sec:RB} we identified one possible explanation for result R1, namely a shift in the power spectrum toward higher wavenumber that leads to enhanced viscous dissipation for a fixed value of $\nu$.  Here we identify two more possible explanations which refer to concisely hereafter as {\bf the Peclet number effect} and {\bf the raising of the adiabat}.   Both of these are linked to the requirement that the convection carry a fixed solar luminosity.  As such, they have no counterpart in classical RB convection where the convective heat flux varies in general with the dissipation (\S\ref{sec:RB}).

In our numerical experiments, as in the Sun, a fixed heat flux enters the convection zone from below by means of radiative diffusion.  We express this as a radiative heat flux $F_r$ that is determined by the heating function $Q$ ($Q$ is equal to the divergence of $F_r$; see eq.\ \ref{eq:Fr}).  This radiative heat flux is the same for each simulation and sets the value of the luminosity that passes through the entire computational domain in equilibrium.

In the mid CZ, the solar luminosity is carried mainly by the convective enthalpy flux, $F_e \propto T v_r$ (see Figs.\ \ref{fig:Fluxbalance} and \ref{fig:fbal_fg} and eq.\ \ref{eq:Fe}).   Here $T$ may be regarded as the temperature deficit of a downward plume relative to its surroundings.  This is because the dominant source of thermal fluctuations is the upper thermal boundary layer (as we will argue in the ensuing discussion in this section).  If these downward plumes are in pressure equilibrium with their surroundings, then $T \sim \overline{T} \delta_p / C_p$, where $\delta_p$ is the entropy deficit in the plume.  In particular, we can set $\delta_p = \left<S\right> - S_p$, where $\left<S\right>$ is the mean entropy gradient and $S_p$ is the specific entropy of a typical cool, downflow plume.  Both $\left<S\right>$ and $S_p$ will in general depend on radius but in what follows we will be concerned with the value of $\delta_p$ in the mid CZ.  Thus, if we equate $v_r$ with $U$ and if require that the convective enthalpy flux must transport the solar luminosity through the mid CZ (neglecting the relatively small contribution from the inward kinetic energy flux), then we must have
\begin{equation}\label{eq:du}
U \propto \frac{L_\odot}{\delta_p} ~~~.
\end{equation}
Since $L_\odot$ is the same for every simulation at all times, $U$ must scale inversely with $\delta_p$.  

Thus, in order to understand results R1, R2, and R3, we must understand the factors that determine $\delta_p$.  To this end, we consider the $\left<S\right>$ curves shown in Fig.\ \ref{fig:Fluxbalance} (right).  Given our fixed entropy upper boundary condition in the C simulations, downward plumes that emanate from the upper boundary layer will be endowed with a specific entropy of $S = 0$ relative to the background state.  If they retain this specific entropy endowment as they travel downward into the mid CZ, then $S_p \approx 0$.  Meanwhile, the background entropy $\left<S\right>$ in the mid CZ is approximately equal to the entropy contrast across the entire CZ, $\Delta S$.   This is due both to the efficient mixing of entropy by the convection, which minimizes $\pd S/\pd r$ apart from the upper boundary layer, and to the lower boundary condition of $\pd S/\pd r = 0$, which can be used to define the base of the CZ\footnote{The base of the CZ may also be defined as the radius where the convective heat flux crosses zero.  With this alternative definition, the point where $\pd S/\pd r = 0$ may not precisely coincide with the base of the CZ but this distinction is irrelevant here.} (note that $\pd S/\pd r = 0$ at the base of the solar CZ as well).

Now the significance of the downflow plumes in determining $\delta_p$ (and $T^\prime$) becomes clear. A plume that begins in the upper boundary layer and travels downward to the mid CZ will have an entropy deficit $\delta_p$ that is comparable to $\Delta S$.  By comparison, fluid flowing up from the lower boundary is nearly isentropic ($S \sim \left<S\right>$) and most of it overturns before it ever reaches the mid CZ as a consequence of the density stratification.  This a characteristic feature of compressible convection \citep{sprui90,miesc05,nordl09}.  

We can generalize this picture by writing $\delta_p = \beta \Delta S$, where $\beta$ is a number between zero and one that quantifies how well downflow plumes can hold on to the specific entropy deficit bequeathed to them in the upper boundary layer.  For example, if $\kappa$ is large, then the thermal content of plumes will diffuse away before they make it to the mid CZ.  This would yield a value of $\beta$ close to zero and inefficient convective heat transport regardless of the value of $\Delta S$.  The relevant nondimensional number here is the Peclet number, which we define here as $P_e = U d_t^2 / (\kappa D)$, where $d_t$ is the width of the thermal boundary layer (equal to the horizontal width of a plume), and $D = r_o - r_i$ is the depth of the CZ.  In the limit $P_e \gg 1$, plumes will be able to retain their thermal variations with negligible diffusive losses.  However, this does not necessarily imply values of $\beta$ close to unity.  Other factors also contribute to $\beta$, most notably the density stratification.  As plumes travel downward, they must entrain surrounding fluid in order to maintain their downward momentum.  This dilutes their entropy content and introduces a multiplicative contribution to $\beta$ that is inversely proportional to the density contrast between the upper and mid CZ.  The value of $\beta$ may also depend on the Reynolds number, which will impact the stability of plumes and may regulate turbulent entrainment.

For our purposes here, we will neglect these additional factors and refer to this phenomena concisely as {\bf the Peclet number effect}.  In short, high values of $P_e$ can promote low velocities $U$ by enhancing the thermal content of plumes, as reflected by an increase in the value of $\beta$.  The contribution of the density stratification to $\beta$ is important in general but not relevant here since all simulations have the same density contrast.

If the SGS heat flux in the surface layers is assumed to be diffusive in nature, $\propto \kappa \pd S / \pd r$ as in \S\ref{sec:Prandtl}, then the Peclet number effect should cease to play a role.  This is because the thermal boundary layer in this case scales as $d_t \propto \kappa^{1/2}$, which would make the diffusion time across a plume, $d_t^2/\kappa$ independent of $\kappa$.  As the diffusion decreases, plumes become smaller, so their diffusive losses are undiminished and their ability to carry heat across the CZ depends only on their speed, $U$.  Lower values of $U$ would then imply lower values of $P_e$, which would in turn lead to lower thermal variations $\delta_p$.  This situation is not consistent with the luminosity constraint (\ref{eq:du}).  However, the Peclet number effect likely does play a role if the width of the boundary layer, $d_t$, is determined by processes other than conduction, as in \S\ref{sec:fg}.  We will return to this issue in \S\ref{sec:ass}.

As noted in \S\ref{sec:numex}, $\Delta S$ increases for the C series of simulations as $\kappa$ is decreased.  This is a consequence of the requirement that conduction must carry the heat flux through the upper boundary, which in turn requires that the upper entropy gradient be $\propto \kappa^{-1}$ (see eq.\ \ref{eq:sgradient}).  Since the width of the thermal boundary layer, $d_t$ scales as $\kappa^{1/2}$, this implies $\Delta S \sim d_t \left[\pd S/\pd r\right]_{r=r_o} \propto \kappa^{-1/2}$.  Given our thermal boundary conditions ($S=0$ at top, $\pd S/\pd r = 0$ at bottom), this effectively raises the specific entropy of the entire CZ, apart from the upper boundary layer.  

We refer to this effect as {\bf the raising of the adiabat}.  In short, a decrease in $\kappa$ will induce a decrease in $U$ by enhancing the thermal deficit of downflow plumes relative to their surroundings, $\delta_p$ (see eq.\ \ref{eq:du}).  It accomplishes this by raising the adiabat of the CZ, enhancing $\Delta S$.

\subsection{High Prandtl Number Convection}\label{sec:highPr}
 
The path in parameter space in which $\nu$ is held fixed while $\kappa$ is decreased (result R1) implies increasingly high values of the Prandtl number, $P_r \gg 1$.  Thermal convection in the regime $P_r \gg$ and $R_a \gg 1$ has been extensively studied in the literature within the context of mantle convection in terrestrial planets.  Notable examples include global simulations of mantle convection that exhibit thin, coherent ``superplumes'' that span the entire convection zone and dominate the heat transport \citep{berco92,desch10,desch12,kronb12}.  Similar structures have also been observed in laboratory experiments \citep{davai07}.  Furthermore, there is some evidence from seismic imaging that such superplumes do in fact exist in the Earth's mantle \citep[e.g.][]{roman02}.

It has been suggested by \cite{sprui97} and \citet{hanas16} that thin, coherent plumes roughly analogous to mantle superplumes may also exist in the solar convection zone, and may help resolve the convective conundrum.  This is an intriguing possibility that warrants further study.  It would be roughly consistent with the scenarios advocated here, in which heat transport is dominated by coherent plumes characterized by high Peclet numbers and thus large thermal contrasts relative to the background (cf. eq.\ \ref{eq:du}).

However, it must be remembered that mantle convection is far different from stellar convection.  In particular, the flow is nearly incompressible and highly viscous, characterized by Reynolds numbers much less than unity (in stars $R_e \gg 1$). Furthermore, large variations in viscosity, thermal conductivity, and the thermal expansion coefficient are all thought to play an important role in the formation of terrestrial superplumes \citep{zhang97,yuen07}.

\subsection{Assessment}\label{sec:ass}

We opened this section (\S\ref{sec:discussion}) with the motivation of understanding result R1; why does $U$ decrease with decreasing $\kappa$ if $\nu$ is held fixed?  In \S\ref{sec:RB}--\ref{sec:lumcon} we offered three potential reasons: (1) a shift in the power spectrum (toward higher wavenumbers), (2) the Peclet number effect, and (3) the raising of the adiabat.  Only the first applies to the well-studied problem of RB convection, the other two arising from the constraint that the convection must carry out the solar luminosity.

\begin{figure}
\begin{center}
\leftline{\includegraphics*[width=0.5\columnwidth]{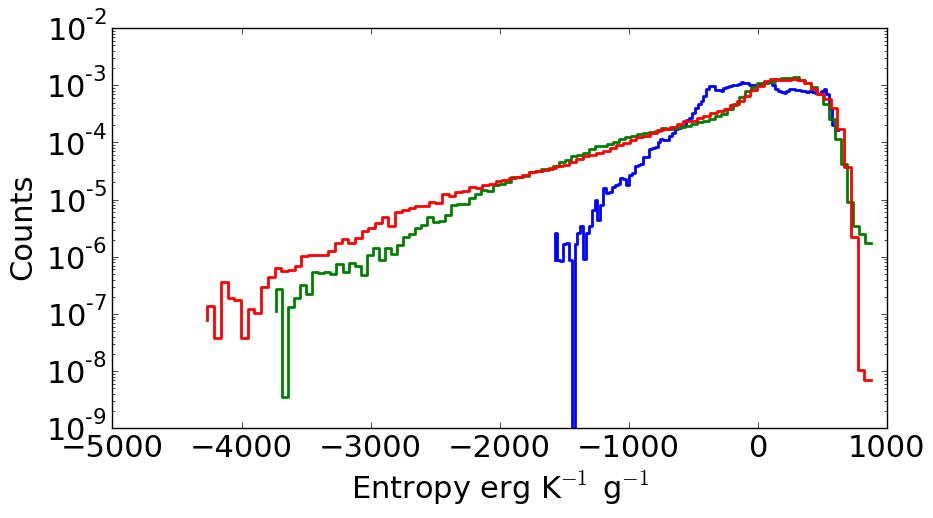}}
\vskip-0.28\columnwidth
\rightline{\includegraphics*[width=0.5\columnwidth]{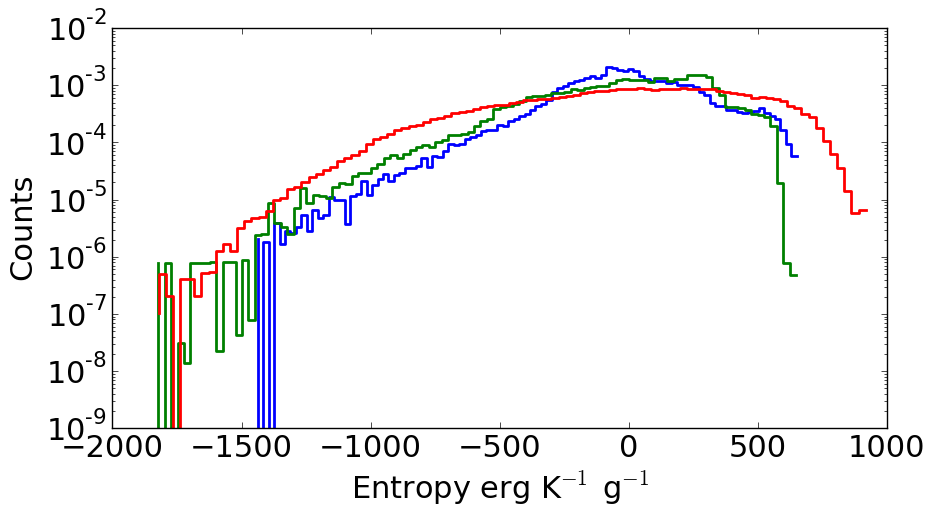}}
\end{center}
\caption{(Left) Probability density function (pdf) of specific entropy variations $S-\left<S\right>$ at $r = 0.83R$ for Cases (blue) C1, (green) C3, and (red) C4.  (Right) As left, but for cases that include the imposed granulation flux $F_g$, namely (blue) G1, (green) G3, and (red) G4.  For those viewing a black and white version, cases C1 and G1 have the shortest negative range, extending only to $\sim -1500$ erg K$^{-1}$ g$^{-1}$ while Cases C4 and G4 extend furthest to the left and right (widest range). \label{fig:spdf}}
\end{figure}

Which effect is dominant?  This is addressed by Figure \ref{fig:spdf}, which shows the distribution of entropy fluctuations in the mid CZ for several cases from section \ref{sec:Prandtl} on the constant-$\nu$ track through parameter space.  This demonstrates an increase in the entropy deficit in cool downflows, $\delta_p$, as $\kappa$ is decreased.  Since (2) the Peclet number effect is expected to be ineffective for these cases with a conductive SGS heat flux (see \S\ref{sec:lumcon}), we attribute the decrease in $U$ with decreasing $\kappa$ (R1) mainly to (3) the raising of the adiabat.

Then what of R2?  Why does the raising of the adiabat become ineffective for the fixed-$P_r$ path through parameter space?  We attribute this to the lower value of $\nu$, which may destabilize plumes and increase turbulent entrainment.  Recall from section \ref{sec:Prandtl} that the effective $R_e$ of the flow increases with decreasing $\kappa$ on this path, in contrast to the fixed-$\nu$ path where the laminarization of plumes (lower $R_e$) may promote more efficient heat transport.  As plumes lose coherence on the fixed-$P_r$ path, the kinetic energy imparted to convection is determined not by the thermal content of individual flow structures, but by the net potential energy available to sustain the convection, which scales as $\Delta S ~ d_t$ (FH16).  As $\Delta S$ increases ($\propto \kappa^{-1/2}$), the thermal boundary layer thins ($d_t \propto \kappa^{1/2}$) so that the potential energy available to sustain convection becomes independent of $\kappa$.  The result is that $U$ becomes independent of dissipation, as was demonstrated by FH16 for $P_r = 1$.  

It may be argued that effects (3) and possibly (1) are artificial in that they depend on our representation of unresolved near-surface convection as a conductive heat flux proportional to $\kappa \pd S/\pd r$.  However, it is clear that our simulations do underestimate $\Delta S$ and over-estimate the thickness of the thermal boundary layer, $d_t$, relative to the actual Sun.  So, we expect that the trends identified here will still be relevant even if the nature of the subgrid-scale heat transport is non-diffusive.

This expectation is borne out by the results presented in \S\ref{sec:fg} (result R3).  Here we partially replaced the conductive SGS heat flux with a fixed heat flux that is independent of both $\kappa$ and $\pd S/\pd r$, in the limit of small $\kappa$.  The results were similar to the conductive case (results R1 and R2).  However, this cannot be attributed to the raising of the adiabat because in these G cases $\Delta S$ does not increase with decreasing $\kappa$ (Fig.\ \ref{fig:fbal_fg}).  Yet, the thermal fluctuations are still enhanced in the low-$\kappa$ limit, as shown in Fig.\ \ref{fig:spdf}.  This points to (2) the Peclet number effect as the likely culprit.  As in the C series, these effects are suppressed in the constant-$P_r$ path through parameter space as the flow becomes more turbulent (higher $R_e$) and plumes lose coherence.

We intend to explore these conclusions more thoroughly in the future with higher-resolution simulations that achieve low enough values of $\kappa$ that the conductive heat flux through the outer surface becomes negligible.  In this regime, effects (1) and (3), namely the shift of the power spectrum and the raising of the adiabat, should cease to play a role because they depend on the structure of the boundary layer, which will become essentially independent of $\kappa$.  However, the Peclet number effect (2) should become more important in this regime, as noted in \S\ref{sec:lumcon}.  This is because the width of the plumes should become independent of $\kappa$ and $P_e$ should scale as $\kappa^{-1}$.

It is worth emphasizing explicitly that all three effects should saturate at very low values of $\kappa$, approaching a regime where $U$ becomes independent of dissipation ($\kappa$ and $\nu$).  For (2) the Peclet number effect, this occurs when $P_e \gg 1$ and thermal diffusion across plumes becomes negligible.  For (1) the shifting of the power spectrum and (3) the raising of the adiabat, saturation should occur when $d_t$ approaches the actual depth of the photospheric boundary layer in the Sun.  Simulations of surface convection suggest that this is on the order of 100 km, if defined as where the buoyancy driving occurs \citep{nordl09}.  However, $\Delta S$ may be established over a slightly wider region, possibly as much as 1 Mm \citep[e.g.][]{lord14}.  This suggests that, $d_t$ is only about 1-2 orders of magnitude thinner than in the highest-resolution simulations considered here.  Thus, extrapolating the scaling relationship $d_t \propto \kappa^{1/2}$ six orders of magnitude or more in $\kappa$ would yield boundary layer widths that are unrealistically thin (FH16).

It is also worth noting that in any asymptotic regime in which $U$ becomes independent of dissipation, then the mean entropy deficit in plumes, $\delta_p$ will also become independent of dissipation.  This follows from eq.\ (\ref{eq:du}).

\section{Summary and Implications}\label{sec:summary}

In this paper we have argued that, if solar convection can operate in a high Prandtl number regime, then this may help alleviate the convective conundrum (Sec.\ \ref{sec:intro}).  This has also been argued by \cite{hotta15b}.  

In particular, we have conducted a series of non-rotating, global solar convection simulations with progressively smaller values of the thermal diffusion, $\kappa$, in order to approach the solar parameter regime.  If $\nu$ is held fixed as we decrease $\kappa$ (corresponding to $P_r > 1$), we find that the amplitude of the convective velocity $U$ systematically decreases (results R1, R3 in \S\ref{sec:discussion}).  For the cases considered in Section \ref{sec:Prandtl} ($F_g = 0$), we attribute this to an increase in the entropy contrast across the CZ, $\Delta S$, which enhances the entropy deficit in cool, downflow plumes.  Larger entropy variations in turn implies that the convection can carry the imposed solar luminosity with smaller flow speeds.  We expect this behavior to saturate when the true $\Delta S$ of the solar convection zone is reached, after which $U$ should become independent of $\kappa$.

An increase in the effective Peclet number can also help reduce $U$ by enhancing thermal fluctuations in the mid CZ.  However, we argue that this should only play a role if the modeled SGS heat flux associated with small-scale surface convection is non-diffusive (\S\ref{sec:ass}).  For a diffusive (conductive) heat flux, the width of plumes becomes narrower as $\kappa$ is decreased so diffusive losses remain undiminished as the thermal diffusivity is reduced.  For a non-diffusive SGS heat flux, we expect that $U$ will become independent of $\kappa$ when $P_e \gg 1$.  These mechanisms for regulating $U$ are distinct from the well-studied problem of Rayleigh-B\'enard convection because they are intimately linked to the requirement that the convection must carry out the solar luminosity (\S\ref{sec:lumcon}).

We find that if we decrease $\nu$ as we decrease $\kappa$ (fixing $P_r$), the results are less favorable from the perspective of the convection conundrum, but still promising.  In particular, as $\kappa$ is decreased, $U$ asymptotes to a constant value that is independent of $\kappa$ (result R2 of \S\ref{sec:discussion}), as found in the more comprehensive study of FH16.

We emphasize again that the simulations reported here are non-rotating.  Further work needs to be done to determine whether our results persist for simulations that include rotation or whether other mechanisms for regulating $U$ might be present.  If they do persist, this raises the question of: Why might solar convection operate in a high $P_r$ regime? 

A plausible answer is small-scale magnetism.  \cite{hotta15b} have recently shown that magnetic fields generated by small-scale dynamo action in high-resolution solar convection simulations can simultaneously inhibit shear and reduce turbulent mixing of entropy fluctuations in downflow plumes.  In other words, the effective viscosity $\nu$ is enhanced and the effective thermal diffusion $\kappa$ is reduced, implying an effective value of $P_r$ that is greater than one.  Furthermore, as the resolution is increased, \cite{hotta15b} argue that the small-scale magnetic energy should saturate, approaching a value that is close to the equipartition value associated with the kinetic energy of the convection.  Thus, one might expect the effective $\nu$ to remain constant as the explicit SGS thermal diffusion (represented by $\kappa$) is decreased.  This provides some justification for our fixed-$\nu$ path through parameter space.

The conclusion that small-scale magnetism can act as an effective viscosity is also supported by a number of global convection simulations \citep{brun04,nelso13,fan14,karak15}.  In particular, \cite{fan14} and \cite{karak15} demonstrate that the suppression of small-scale shear by magnetism can modify the transition from solar to anti-solar differential rotation discussed in Sec.\ \ref{sec:intro}, shifting it toward higher Rossby numbers.  In other words, the inclusion of magnetism can promote solar-like differential rotation in a simulation that would otherwise be anti-solar.

Though promising, these results must still be considered speculative; more work is needed to determine whether or not solar convection effectively operates in the high-$P_r$ regime and if so, whether or not this can solve the convection conundrum. 

We thank Ben Brown, Sacha Brun, Brad Hindman, Hideyuki Hotta, Bidya Karak, Mark Rast, and Matthias Rempel for many enlightening discussions on the issues covered here and we thank the anonymous referees for helpful comments on the manuscript.  This research is supported by NASA grant NNX13AG18G.  O'Mara was supported by the Research Experience for Undergraduates (REU) program at University of Colorado's Laboratory for Atmospheric and Space Physics (LASP) and by a research stipend from Regis University.  Featherstone and the development of {\em Rayleigh} were supported by the Computational Infrastructure for Geodynamics (CIG), which is supported by the National Science Foundation award NSF-094946.  This work used computational resources provided through NASA HEC support on the Pleiades supercomputer in conjunction with NASA grant NNX14AC05G.  NCAR is sponsored by the National Science Foundation.



\begin{thebibliography}{31}
\expandafter\ifx\csname natexlab\endcsname\relax\def\natexlab#1{#1}\fi
\providecommand{\url}[1]{\texttt{#1}}
\providecommand{\href}[2]{#2}
\providecommand{\path}[1]{#1}
\providecommand{\DOIprefix}{doi:}
\providecommand{\ArXivprefix}{arXiv:}
\providecommand{\URLprefix}{URL: }
\providecommand{\Pubmedprefix}{pmid:}
\providecommand{\doi}[1]{\href{http://dx.doi.org/#1}{\path{#1}}}
\providecommand{\Pubmed}[1]{\href{pmid:#1}{\path{#1}}}
\providecommand{\bibinfo}[2]{#2}
\ifx\xfnm\relax \def\xfnm[#1]{\unskip,\space#1}\fi
\bibitem[{Ahlers et~al.(2009)Ahlers, Grossmann and Lohse}]{ahler09}
\bibinfo{author}{Ahlers, G.}, \bibinfo{author}{Grossmann, S.},
  \bibinfo{author}{Lohse, D.}, \bibinfo{year}{2009}.
\newblock \bibinfo{title}{Heat transfer and large scale dynamics in turbulent
  rayleigh-b\'enard convection}.
\newblock \bibinfo{journal}{Reviews of Modern Physics} \bibinfo{volume}{81},
  \bibinfo{pages}{503--537}.
\bibitem[{Augustson et~al.(2015)Augustson, Brun, Miesch and Toomre}]{augus15}
\bibinfo{author}{Augustson, K.C.}, \bibinfo{author}{Brun, A.S.},
  \bibinfo{author}{Miesch, M.S.}, \bibinfo{author}{Toomre, J.},
  \bibinfo{year}{2015}.
\newblock \bibinfo{title}{Grand minima and equatorward propagation in a cycling
  stellar convective dynamo}.
\newblock \bibinfo{journal}{ApJ} \bibinfo{volume}{809}, \bibinfo{pages}{149
  (25pp)}.
\bibitem[{Bercovici et~al.(1992)Bercovici, Schubert and Glatzmaier}]{berco92}
\bibinfo{author}{Bercovici, D.}, \bibinfo{author}{Schubert, G.},
  \bibinfo{author}{Glatzmaier, G.A.}, \bibinfo{year}{1992}.
\newblock \bibinfo{title}{Three-dimensional convection of an
  infinite-prandtl-number compressible fluid in a basally heated spherical
  shell}.
\newblock \bibinfo{journal}{J.\ Fluid Mech.} \bibinfo{volume}{239},
  \bibinfo{pages}{683--719}.
\bibitem[{Brun et~al.(2004)Brun, Miesch and Toomre}]{brun04}
\bibinfo{author}{Brun, A.S.}, \bibinfo{author}{Miesch, M.S.},
  \bibinfo{author}{Toomre, J.}, \bibinfo{year}{2004}.
\newblock \bibinfo{title}{Global-scale turbulent convection and magnetic dynamo
  action in the solar envelope}.
\newblock \bibinfo{journal}{ApJ} \bibinfo{volume}{614},
  \bibinfo{pages}{1073--1098}.
\bibitem[{Chandrasekhar(1961)}]{chand61}
\bibinfo{author}{Chandrasekhar, S.}, \bibinfo{year}{1961}.
\newblock \bibinfo{title}{Hydrodynamic and Hydromagnetic Stability}.
\newblock \bibinfo{publisher}{Oxford University Press},
  \bibinfo{address}{Oxford}.
\bibitem[{Chapman and Proctor(1980)}]{chapm80}
\bibinfo{author}{Chapman, C.J.}, \bibinfo{author}{Proctor, M.R.E.},
  \bibinfo{year}{1980}.
\newblock \bibinfo{title}{Nonlinear rayleigh b\'enard convection between poorly
  conducting boundaries}.
\newblock \bibinfo{journal}{J.\ Fluid Mech.} \bibinfo{volume}{101},
  \bibinfo{pages}{759--782}.
\bibitem[{Charbonneau(2010)}]{charb10}
\bibinfo{author}{Charbonneau, P.}, \bibinfo{year}{2010}.
\newblock \bibinfo{title}{Dynamo models of the solar cycle}.
\newblock \bibinfo{journal}{Living Reviews in Solar Physics}
  \bibinfo{volume}{7}.
\newblock \bibinfo{note}{Http://www.livingreviews.org/lrsp-2010-3}.
\bibitem[{Clune et~al.(1999)Clune, Elliott, Miesch, Toomre and
  Glatzmaier}]{clune99}
\bibinfo{author}{Clune, T.C.}, \bibinfo{author}{Elliott, J.R.},
  \bibinfo{author}{Miesch, M.S.}, \bibinfo{author}{Toomre, J.},
  \bibinfo{author}{Glatzmaier, G.A.}, \bibinfo{year}{1999}.
\newblock \bibinfo{title}{Computational aspects of a code to study rotating
  turbulent convection in spherical shells}.
\newblock \bibinfo{journal}{Parallel Computing} \bibinfo{volume}{25},
  \bibinfo{pages}{361--380}.
\bibitem[{Davaille and Limare(2007)}]{davai07}
\bibinfo{author}{Davaille, A.}, \bibinfo{author}{Limare, A.},
  \bibinfo{year}{2007}.
\newblock \bibinfo{title}{Laboratory Studies of Mantle Convection}.
  \bibinfo{booktitle}{Treatise on Geophysics, Volume 7: Mantle Dynamics, second edition},
  chapter~\bibinfo{chapter}{3}.
  \bibinfo{publisher}{Elsevier}, \bibinfo{address}{Amsterdam}. pp. 89--165
\bibitem[{Depassier and Spiegel(1981)}]{depas81}
\bibinfo{author}{Depassier, M.C.}, \bibinfo{author}{Spiegel, E.A.},
  \bibinfo{year}{1981}.
\newblock \bibinfo{title}{The large-scale structure of compressible
  convection}.
\newblock \bibinfo{journal}{Astronom.\ J.} \bibinfo{volume}{86},
  \bibinfo{pages}{496--512}.
\bibitem[{Deschamps et~al.(2010)Deschamps, Tackley and Nakagawa}]{desch10}
\bibinfo{author}{Deschamps, F.}, \bibinfo{author}{Tackley, P.J.},
  \bibinfo{author}{Nakagawa, T.}, \bibinfo{year}{2010}.
\newblock \bibinfo{title}{Temperature and heat flux scalings for isoviscous
  thermal convection in spherical geometry}.
\newblock \bibinfo{journal}{Geophys.\ J.\ Int.} \bibinfo{volume}{182},
  \bibinfo{pages}{137--154}.
\bibitem[{Deschamps et~al.(2012)Deschamps, Yao, Tackley and
  Sanchez-Valle}]{desch12}
\bibinfo{author}{Deschamps, F.}, \bibinfo{author}{Yao, C.},
  \bibinfo{author}{Tackley, P.J.}, \bibinfo{author}{Sanchez-Valle, C.},
  \bibinfo{year}{2012}.
\newblock \bibinfo{title}{High rayleigh number thermal convection in
  volumetrically heated spherical shells}.
\newblock \bibinfo{journal}{J.\ Geophys.\ Res.} \bibinfo{volume}{117},
  \bibinfo{pages}{E09006}.
\bibitem[{Fan and Fang(2014)}]{fan14}
\bibinfo{author}{Fan, Y.}, \bibinfo{author}{Fang, F.}, \bibinfo{year}{2014}.
\newblock \bibinfo{title}{A simulation of convective dynamo in the solar
  convective envelope: Maintenance of the solar-like differential rotation and
  emerging flux}.
\newblock \bibinfo{journal}{ApJ} \bibinfo{volume}{789}, \bibinfo{pages}{35
  (13pp)}.
\bibitem[{Featherstone and Hindman(2016)}]{feath16}
\bibinfo{author}{Featherstone, N.A.}, \bibinfo{author}{Hindman, B.W.},
  \bibinfo{year}{2016}.
\newblock \bibinfo{title}{The spectral amplitude of stellar convection and its
  scaling in the high-rayleigh-number regime}.
\newblock \bibinfo{note}{Astrophys.\ J., submitted}.
\bibitem[{Featherstone and Miesch(2015)}]{feath15}
\bibinfo{author}{Featherstone, N.A.}, \bibinfo{author}{Miesch, M.S.},
  \bibinfo{year}{2015}.
\newblock \bibinfo{title}{Meridional circulation in solar and stellar
  convection zones}.
\newblock \bibinfo{journal}{ApJ} \bibinfo{volume}{804}, \bibinfo{pages}{67
  (21pp)}.
\bibitem[{Gastine et~al.(2015)Gastine, Wicht and Aurnou}]{gasti15}
\bibinfo{author}{Gastine, T.}, \bibinfo{author}{Wicht, J.},
  \bibinfo{author}{Aurnou, J.M.}, \bibinfo{year}{2015}.
\newblock \bibinfo{title}{Turbulent rayleigh-b\'enard convection in spherical
  shells}.
\newblock \bibinfo{journal}{J.\ Fluid Mech.} \bibinfo{volume}{778},
  \bibinfo{pages}{721--764}.
\bibitem[{Gastine et~al.(2014)Gastine, Yadav, Morin, Reiners and
  Wicht}]{gasti14}
\bibinfo{author}{Gastine, T.}, \bibinfo{author}{Yadav, R.K.},
  \bibinfo{author}{Morin, J.}, \bibinfo{author}{Reiners, A.},
  \bibinfo{author}{Wicht, J.}, \bibinfo{year}{2014}.
\newblock \bibinfo{title}{From solar-like to anti-solar differential rotation
  in cool stars}.
\newblock \bibinfo{journal}{MNRAS} \bibinfo{volume}{438},
  \bibinfo{pages}{L76--L80}.
\bibitem[{Gilman(1983)}]{gilma83}
\bibinfo{author}{Gilman, P.A.}, \bibinfo{year}{1983}.
\newblock \bibinfo{title}{Dynamically consistent nonlinear dynamos driven by
  convection in a rotating spherical shell ii. dynamos with cycles and strong
  feedbacks}.
\newblock \bibinfo{journal}{ApJSS} \bibinfo{volume}{53},
  \bibinfo{pages}{243--268}.
\bibitem[{Gilman and Glatzmaier(1981)}]{gilma81b}
\bibinfo{author}{Gilman, P.A.}, \bibinfo{author}{Glatzmaier, G.A.},
  \bibinfo{year}{1981}.
\newblock \bibinfo{title}{Compressible convection in a rotating spherical
  shell. i. anelastic equations}.
\newblock \bibinfo{journal}{ApJSS} \bibinfo{volume}{45},
  \bibinfo{pages}{335--349}.
\bibitem[{Glatzmaier(1985)}]{glatz85a}
\bibinfo{author}{Glatzmaier, G.A.}, \bibinfo{year}{1985}.
\newblock \bibinfo{title}{Numerical simulations of stellar convective dynamos.
  ii. field propogation in the convection zone}.
\newblock \bibinfo{journal}{ApJ} \bibinfo{volume}{291},
  \bibinfo{pages}{300--307}.
\bibitem[{Gough(1969)}]{gough69}
\bibinfo{author}{Gough, D.O.}, \bibinfo{year}{1969}.
\newblock \bibinfo{title}{The anelastic approximation for thermal convection}.
\newblock \bibinfo{journal}{J. Atmos. Sci.} \bibinfo{volume}{26},
  \bibinfo{pages}{448--456}.
\bibitem[{Greer et~al.(2015)Greer, Hindman, Featherstone and Toomre}]{greer15}
\bibinfo{author}{Greer, B.J.}, \bibinfo{author}{Hindman, B.W.},
  \bibinfo{author}{Featherstone, N.A.}, \bibinfo{author}{Toomre, J.},
  \bibinfo{year}{2015}.
\newblock \bibinfo{title}{Helioseismic imaging of fast convective flows
  throughout the near-surface shear layer}.
\newblock \bibinfo{journal}{ApJ} \bibinfo{volume}{803}, \bibinfo{pages}{L17
  (5pp)}.
\bibitem[{Guerrero et~al.(2013)Guerrero, Smolarkiewicz, Kosovichev and
  Mansour}]{guerr13}
\bibinfo{author}{Guerrero, G.}, \bibinfo{author}{Smolarkiewicz, P.K.},
  \bibinfo{author}{Kosovichev, A.G.}, \bibinfo{author}{Mansour, N.N.},
  \bibinfo{year}{2013}.
\newblock \bibinfo{title}{Differential rotation in solar-like stars from global
  simulations}.
\newblock \bibinfo{journal}{ApJ} \bibinfo{volume}{779}, \bibinfo{pages}{176
  (13pp)}.
\bibitem[{Hanasoge et~al.(2010)Hanasoge, Duvall and DeRosa}]{hanas10}
\bibinfo{author}{Hanasoge, S.M.}, \bibinfo{author}{Duvall, T.},
  \bibinfo{author}{DeRosa, M.L.}, \bibinfo{year}{2010}.
\newblock \bibinfo{title}{Seismic constraints on interior solar convection}.
\newblock \bibinfo{journal}{ApJL} \bibinfo{volume}{712},
  \bibinfo{pages}{L98--L102}.
\bibitem[{Hanasoge et~al.(2012)Hanasoge, Duvall and Sreenivasan}]{hanas12}
\bibinfo{author}{Hanasoge, S.M.}, \bibinfo{author}{Duvall, T.},
  \bibinfo{author}{Sreenivasan, K.R.}, \bibinfo{year}{2012}.
\newblock \bibinfo{title}{Anamolously weak solar convection}.
\newblock \bibinfo{journal}{Proc.\ Nat.\ Acad.\ Sci.} \bibinfo{volume}{109},
  \bibinfo{pages}{11928--11932}, \newblock \bibinfo{note}{doi:10.1073/pnas.1206570109}.
\bibitem[{Hanasoge et~al.(2016)Hanasoge, Gizon and Sreenivasan}]{hanas16}
\bibinfo{author}{Hanasoge, S.M.}, \bibinfo{author}{Gizon, L.},
  \bibinfo{author}{Sreenivasan, K.R.}, \bibinfo{year}{2016}.
\newblock \bibinfo{title}{Seismic sounding of convection in the sun}.
\newblock \bibinfo{journal}{Ann.\ Rev.\ Fluid Mech.} \bibinfo{volume}{48},
  \bibinfo{pages}{191--217}.
\bibitem[{Hotta et~al.(2014)Hotta, Rempel and Yokoyama}]{hotta14a}
\bibinfo{author}{Hotta, H.}, \bibinfo{author}{Rempel, M.},
  \bibinfo{author}{Yokoyama, T.}, \bibinfo{year}{2014}.
\newblock \bibinfo{title}{High-resolution calculations of the solar global
  convection with the reduced speed of sound technique. i. the structure of the
  convection and the magnetic field without the rotation}.
\newblock \bibinfo{journal}{ApJ} \bibinfo{volume}{786}, \bibinfo{pages}{24
  (18pp)}.
\bibitem[{Hotta et~al.(2015a)Hotta, Rempel and Yokoyama}]{hotta15a}
\bibinfo{author}{Hotta, H.}, \bibinfo{author}{Rempel, M.},
  \bibinfo{author}{Yokoyama, T.}, \bibinfo{year}{2015}a.
\newblock \bibinfo{title}{High-resolution calculations of the solar global
  convection with the reduced speed of sound technique. ii. near surface shear
  layer with the rotation}.
\newblock \bibinfo{journal}{ApJ} \bibinfo{volume}{798}, \bibinfo{pages}{51 (15pp)}.
\bibitem[{Hotta et~al.(2015b)Hotta, Rempel and Yokoyama}]{hotta15b}
\bibinfo{author}{Hotta, H.}, \bibinfo{author}{Rempel, M.},
  \bibinfo{author}{Yokoyama, T.}, \bibinfo{year}{2015}b.
\newblock \bibinfo{title}{Efficient small-scale dynamo action in the solar
  convection zone}.
\newblock \bibinfo{journal}{ApJ} \bibinfo{volume}{803}, \bibinfo{pages}{42 (14pp)}.
\bibitem[{Jones et~al.(2011)Jones, Boronski, Brun, Glatzmaier, Gastine, Miesch
  and Wicht}]{jones11}
\bibinfo{author}{Jones, C.A.}, \bibinfo{author}{Boronski, P.},
  \bibinfo{author}{Brun, A.S.}, \bibinfo{author}{Glatzmaier, G.A.},
  \bibinfo{author}{Gastine, T.}, \bibinfo{author}{Miesch, M.S.},
  \bibinfo{author}{Wicht, J.}, \bibinfo{year}{2011}.
\newblock \bibinfo{title}{Anelastic convection-driven dynamo benchmarks}.
\newblock \bibinfo{journal}{Icarus} \bibinfo{volume}{216},
  \bibinfo{pages}{120--135}.
\bibitem[{K\"apyl\"a et~al.(2014)K\"apyl\"a, K\"apyl\"a and Brandenburg}]{kapyl14}
\bibinfo{author}{K\"apyl\"a, P.J.}, \bibinfo{author}{K\"apyl\"a, M.J.},
  \bibinfo{author}{Brandenburg, A.}, \bibinfo{year}{2014}.
\newblock \bibinfo{title}{Confirmation of bistable stellar differential
  rotation profiles}.
\newblock \bibinfo{journal}{A\&A} \bibinfo{volume}{570}, \bibinfo{pages}{A43 (10pp)}.
\bibitem[{Karak et~al.(2015)Karak, K\"apyl\"a, K\"apyl\"a, Brandenburg, Olspert
  and Pelt}]{karak15}
\bibinfo{author}{Karak, B.B.}, \bibinfo{author}{K\"apyl\"a, P.J.},
  \bibinfo{author}{K\"apyl\"a, M.J.}, \bibinfo{author}{Brandenburg, A.},
  \bibinfo{author}{Olspert, N.}, \bibinfo{author}{Pelt, J.},
  \bibinfo{year}{2015}.
\newblock \bibinfo{title}{Magnetically controlled stellar differential rotation
  near the transition from solar to anti-solar profiles}.
\newblock \bibinfo{journal}{A\&A} \bibinfo{volume}{576}, \bibinfo{pages}{A26
  (17pp)}.
\bibitem[{Kronbichler et~al.(2012)Kronbichler, Heister and Bangerth}]{kronb12}
\bibinfo{author}{Kronbichler, M.}, \bibinfo{author}{Heister, T.},
  \bibinfo{author}{Bangerth, W.}, \bibinfo{year}{2012}.
\newblock \bibinfo{title}{High accuracy mantle convection simulation through
  modern numerical methods}.
\newblock \bibinfo{journal}{Geophys.\ J.\ Int.} \bibinfo{volume}{191},
  \bibinfo{pages}{12--29}.
\bibitem[{Lord et~al.(2014)Lord, Cameron, Rast, Rempel and Roudier}]{lord14}
\bibinfo{author}{Lord, J.}, \bibinfo{author}{Cameron, R.},
  \bibinfo{author}{Rast, M.}, \bibinfo{author}{Rempel, M.},
  \bibinfo{author}{Roudier, T.}, \bibinfo{year}{2014}.
\newblock \bibinfo{title}{The role of subsurface flows in solar surface
  convection: Modeling the spectrum of supergranular and larger scale flows}.
\newblock \bibinfo{journal}{ApJ} \bibinfo{volume}{793}, \bibinfo{pages}{24
  (11pp)}.
\bibitem[{Miesch(2005)}]{miesc05}
\bibinfo{author}{Miesch, M.S.}, \bibinfo{year}{2005}.
\newblock \bibinfo{title}{Large-scale dynamics of the convection zone and
  tachocline}.
\newblock \bibinfo{journal}{Living Reviews in Solar Physics}
  \bibinfo{volume}{2}.
\newblock \bibinfo{note}{Http://www.livingreviews.org/lrsp-2005-1}.
\bibitem[{Miesch et~al.(2012)Miesch, Featherstone, Rempel and
  Trampedach}]{miesc12b}
\bibinfo{author}{Miesch, M.S.}, \bibinfo{author}{Featherstone, N.A.},
  \bibinfo{author}{Rempel, M.}, \bibinfo{author}{Trampedach, R.},
  \bibinfo{year}{2012}.
\newblock \bibinfo{title}{On the amplitude of convective velocities in the deep
  solar interior}.
\newblock \bibinfo{journal}{ApJ} \bibinfo{volume}{757}, \bibinfo{pages}{128
  (14pp)}.
\bibitem[{Nelson et~al.(2013)Nelson, Brown, Brun, Miesch and Toomre}]{nelso13}
\bibinfo{author}{Nelson, N.J.}, \bibinfo{author}{Brown, B.P.},
  \bibinfo{author}{Brun, A.S.}, \bibinfo{author}{Miesch, M.S.},
  \bibinfo{author}{Toomre, J.}, \bibinfo{year}{2013}.
\newblock \bibinfo{title}{Magnetic wreathes and cycles in convective dynamos}.
\newblock \bibinfo{journal}{ApJ} \bibinfo{volume}{762}, \bibinfo{pages}{73
  (20pp)}.
\bibitem[{Nordlund et~al.(2009)Nordlund, Stein and Asplund}]{nordl09}
\bibinfo{author}{Nordlund, A.}, \bibinfo{author}{Stein, R.F.},
  \bibinfo{author}{Asplund, M.}, \bibinfo{year}{2009}.
\newblock \bibinfo{title}{Solar surface convection}.
\newblock \bibinfo{journal}{Living Reviews in Solar Physics}
  \bibinfo{volume}{6}.
\newblock \bibinfo{note}{Http://www.livingreviews.org/lrsp-2009-2}.
\bibitem[{Racine et~al.(2011)Racine, Charbonneau, Ghizaru, Bouchat and
  Smolarkiewicz}]{racin11}
\bibinfo{author}{Racine, E.}, \bibinfo{author}{Charbonneau, P.},
  \bibinfo{author}{Ghizaru, M.}, \bibinfo{author}{Bouchat, A.},
  \bibinfo{author}{Smolarkiewicz, P.K.}, \bibinfo{year}{2011}.
\newblock \bibinfo{title}{On the mode of dynamo action in a global large-eddy
  simulation of solar convection}.
\newblock \bibinfo{journal}{ApJ} \bibinfo{volume}{735}, \bibinfo{pages}{46
  (22pp)}.
\bibitem[{Romanowicz and Gung(2002)}]{roman02}
\bibinfo{author}{Romanowicz, B.}, \bibinfo{author}{Gung, Y.},
  \bibinfo{year}{2002}.
\newblock \bibinfo{title}{Superplumes from the core-mantle boundary to the
  lithosphere: Implications for heat flux}.
\newblock \bibinfo{journal}{Science} \bibinfo{volume}{296},
  \bibinfo{pages}{513--516}.
\bibitem[{Siggia(1994)}]{siggi94}
\bibinfo{author}{Siggia, E.D.}, \bibinfo{year}{1994}.
\newblock \bibinfo{title}{High rayleigh number convection}.
\newblock \bibinfo{journal}{Annu. Rev. Fluid Mech.} \bibinfo{volume}{26},
  \bibinfo{pages}{137--168}.
\bibitem[{Spruit(1997)}]{sprui97}
\bibinfo{author}{Spruit, H.C.}, \bibinfo{year}{1997}.
\newblock \bibinfo{title}{Convection in stellar envelopes: A changing
  paradigm}.
\newblock \bibinfo{journal}{Mem. Soc. Astr. It.} \bibinfo{volume}{68},
  \bibinfo{pages}{397--414}.
\bibitem[{Spruit et~al.(1990)Spruit, Nordlund and Title}]{sprui90}
\bibinfo{author}{Spruit, H.C.}, \bibinfo{author}{Nordlund, A.},
  \bibinfo{author}{Title, A.M.}, \bibinfo{year}{1990}.
\newblock \bibinfo{title}{Solar convection}.
\newblock \bibinfo{journal}{Annu. Rev. Astron. Astrophys.}
  \bibinfo{volume}{28}, \bibinfo{pages}{263--301}.
\bibitem[{Yuen et~al.(2007)Yuen, Monnereau, Hansen, Kameyama and
  Matyska}]{yuen07}
\bibinfo{author}{Yuen, D.A.}, \bibinfo{author}{Monnereau, M.},
  \bibinfo{author}{Hansen, U.}, \bibinfo{author}{Kameyama, M.},
  \bibinfo{author}{Matyska, C.}, \bibinfo{year}{2007}.
\newblock \bibinfo{title}{Dynamics of superplumes in the lower mantle}, in:
  \bibinfo{editor}{Yuen, D.A.}, \bibinfo{editor}{Maruyama, S.},
  \bibinfo{editor}{Karato, S.I.}, \bibinfo{editor}{Windley, B.F.} (Eds.),
  \bibinfo{booktitle}{Superplumes: Beyond Plate Tectonics},
  \bibinfo{publisher}{Springer}, \bibinfo{address}{Dordrecht}. pp.
  \bibinfo{pages}{239--267}.
\bibitem[{Zhang et~al.(1997)Zhang, Childress and Libchaber}]{zhang97}
\bibinfo{author}{Zhang, J.}, \bibinfo{author}{Childress, S.},
  \bibinfo{author}{Libchaber, A.}, \bibinfo{year}{1997}.
\newblock \bibinfo{title}{Non-boussinesq effect: Thermal convection with broken
  symmetry}.
\newblock \bibinfo{journal}{Phys.\ Fluids} \bibinfo{volume}{9},
  \bibinfo{pages}{1034--1042}.

\end{thebibliography}

\end{document}